\documentclass[aps,pra,amsmath,amssymb,reprint]{revtex4-2}
\usepackage{graphicx}
\usepackage{bm}
\begin{document}
%
%
\preprint{ETRI 2025-000} 
%

%
\title{An optical transistor of the nonlinear resonant structure}
%


%
\author{Jongbae Kim}
\email[]{jongbae@etri.re.kr}
\email[]{jongbae@sci.kr}
\affiliation{Quantum Technology Research Division, ETRI, Daejeon 34129, Republic of Korea \\
KREONET Center, KISTI, Daejeon 34141, Republic of Korea}
%
\date{arXiv, \today}
\begin{abstract}
An optical transistor capable of simultaneous amplification and switching  
is theoretically proposed 
via  cascaded second-order nonlinear interactions  
in a resonant structure.  
Two distinct operational schemes are analyzed. 
A single frequency scheme employs cascaded second harmonic generation  
and inverse second harmonic generation (SHG/iSHG) 
using two Type-I SHG interactions, 
whereas a dual frequency scheme employs cascaded 
SHG and optical parametric amplification (SHG/OPA). 
Exact theoretical solutions and numerical calculations 
show cascadable amplification and digital on/off switching.  
%
A new optical phenomenon of nonlinear transparency is predicted
by the theoretical solutions and confirmed by the numerical solutions 
in each scheme of the cascaded SHG/iSHG and SHG/OPA.
The single and dual frequency configurations satisfy the cascadability 
and fan-out criteria with power transfer ratios of 
$\alpha_{TR}=4.838$ and $52.26$ and power amplification factors of 
$\beta_{AF}=48.38$ and $522.6$, respectively.
%
These results indicate transistor-like performance  at input powers 
in the milliwatt range, readily supplied by laser diodes.  
The proposed structure establishes a physically feasible 
and practically scalable route to optical transistors 
operating at  high speed and low power for integrated photonic circuits,
with broad applications in all optical communication and computing.
\end{abstract}


%
\maketitle
%

%

\section{\label{Intro} Introduction}
%
An optical transistor is a device 
that amplifies and switches optical signals simultaneously. 
If the optical transistor can be implemented, 
all optical processing will be possible in various fields of optical technologies, 
especially enabling the implementation of optical computing 
with high-speed performance beyond that of electronic computing.
However, 
since optical waves do not interact with each other in classical linear optics
due to the zero charge of a photon, which cannot be controlled by electric potential, 
it is difficult to control one optical wave with another for all optical processing. 
Despite many difficulties, 
various schemes using multiple media have been proposed 
to implement optical transistors. 
These include 
silicon microring configuration \cite{Varghese2012}, 
electromagnetically induced absorption 
using a microresonator \cite{Clader2013}, 
a one-dimensional photonic-crystal cavity 
with an active Raman gain medium \cite{Arkhipkin2013}, 
polariton fluids in the plane of the microcavity \cite{Ballarini2013}, 
optical control of exciton fluxes \cite{Andreakou2014}, 
tunneling of light through frustrated total internal reflection 
in a photonic crystal structure \cite{Goodarzi2018}, 
a microcavity with a ladder-type polymer 
for polariton relaxation \cite{Zasedatelev2019},
deep-level defects in nitride semiconductors \cite{Matys2020}, and
single-photon nonlinearity in a polymer-coupled microcavity 
forming exciton-polaritons \cite{Zasedatelev2021}.
Meanwhile, since the observation of SHG \cite{Franken1961} and
the theoretical establishment 
of the nonlinear optical interactions \cite{Armstrong1962}, 
based on the characteristics that induce direct interactions among optical waves,
second-order nonlinear interactions have been applied to the research 
on the optical transistor.
This research includes
a device wherein a weak incoming signal controls 
a strong outgoing signal utilizing SHG \cite{Jain1976} and
frequency-degenerate quadratic three-wave interactions 
in a Type-II SHG geometry \cite{S.Kim1998}. 
However, it is still known that an optical transistor 
comparable to  an electronic transistor has not  been realized yet. 
%

%
 In spite of extensive efforts to realize an optical transistor 
 using diverse media and device schemes 
 in Refs.~[1--13], a practical implementation is still elusive.
An optical transistor should satisfy stringent criteria, 
particularly cascadability and fan-out, 
in order to be practically useful \cite{Miller2010}. 
For comparison, an electronic transistor normally exhibits 
a current transfer ratio 
$\alpha \simeq 0.95\!-\!0.99$ 
and a current amplification factor 
$\beta \simeq 20\!-\!200$, 
while an analogous optical device should achieve a power transfer ratio 
$\alpha_{TR}  \geq 2$ for the fan-out
and a power amplification factor $\beta_{AF} \approx \beta $
to be comparable to electronic transistors.
Several proposals in Refs.~[1--11] have demonstrated simultaneous 
amplification and switching; 
however, realizing a practical optical transistor that satisfies 
the demanding criteria and achieves performance truly comparable 
to that of an electronic transistor remains a significant challenge.
Aside from the schemes in Refs.~[1--11], the two approaches described 
in Refs.~\cite{Jain1976, S.Kim1998} are based on 
classical second-order nonlinear equations and theoretical analysis; nevertheless, 
they also suffer from fundamental limitations—namely, 
excessively high pump requirements 
($\sim 10^{6}$~W or $\sim 1$~GW/cm$^{2}$), lack of a clear off state, 
and inability to satisfy the cascadability and the fan-out requirements—thereby 
hindering practical feasibility.
In contrast, 
the proposed optical transistor employs a nonlinear waveguide 
within a resonant structure as a single compact device, 
enabling efficient cascaded interactions, 
well defined digital on/off states, and operation at low input powers 
readily supplied by commercial laser diodes.
The collector output exceeds both base and emitter inputs, 
achieving sufficiently large $\alpha_{TR}$  and $\beta_{AF}$,
thereby demonstrating compliance with the cascadability and the
fan-out requirements among the criteria. 
Moreover, the underlying 
cascaded SHG/iSHG and SHG/OPA 
processes have been theoretically analyzed and numerically verified, 
clarifying how physical parameters govern transistor behavior. 
Through these analyses, the proposed structure is shown 
to overcome the limitations of earlier designs 
and to achieve performance levels that are comparable to, 
and in some respects competitive with, electronic transistors.
%
%

%
In this paper, to address the issue
of realizing an optical transistor comparable to an electronic transistor,
two schemes for the optical transistor 
based on the second-order nonlinear interactions 
in the resonant structure are proposed.
For each scheme,  either single or dual frequency optical waves
are needed for the operation.
The single frequency waves interact with each other 
through cascaded SHG/iSHG, 
whereas the dual frequency waves interact with each other 
through cascaded SHG/OPA \cite{Gallo1997}.
In the optical transistor,
the base wave controls the emitter wave and in turn the collector wave, 
thereby implementing the simultaneous amplification 
and switching of the optical signals.
This paper is organized as follows.
In Sec.~\ref{OTR-1},
for one scheme of the optical transistor operating with waves 
of  a single frequency, 
a theory of two Type-I SHG interactions in the nonlinear resonant structure
is proposed.
Theoretical solutions of the cascaded SHG/iSHG and SHG/SHG equations are 
derived for the analysis of  the two Type-I SHG interactions.
 Numerical solutions for the cascaded SHG/iSHG are calculated and plotted
 to demonstrate the amplification and the switching.
In Sec.~\ref{OTR-2},
for the other scheme of the optical transistor operating with waves 
of  dual frequencies,
the cascaded SHG/OPA interactions in the nonlinear resonant structure
are considered.
Theoretical solutions of the cascaded SHG/OPA are derived 
and numerical solutions are calculated
for the analysis and the illustration.
Finally, implications for the optical transistor 
of the nonlinear resonant structure 
are summarized in Sec.~\ref{Conclusions}.
\section{\label{OTR-1} An optical transistor
\\ 
with waves of a single frequency}  
\subsection{\label{OTR-1-theory} Theory of two Type-I SHGs}
\begin{figure}
\includegraphics[width=8.0cm, angle=0]{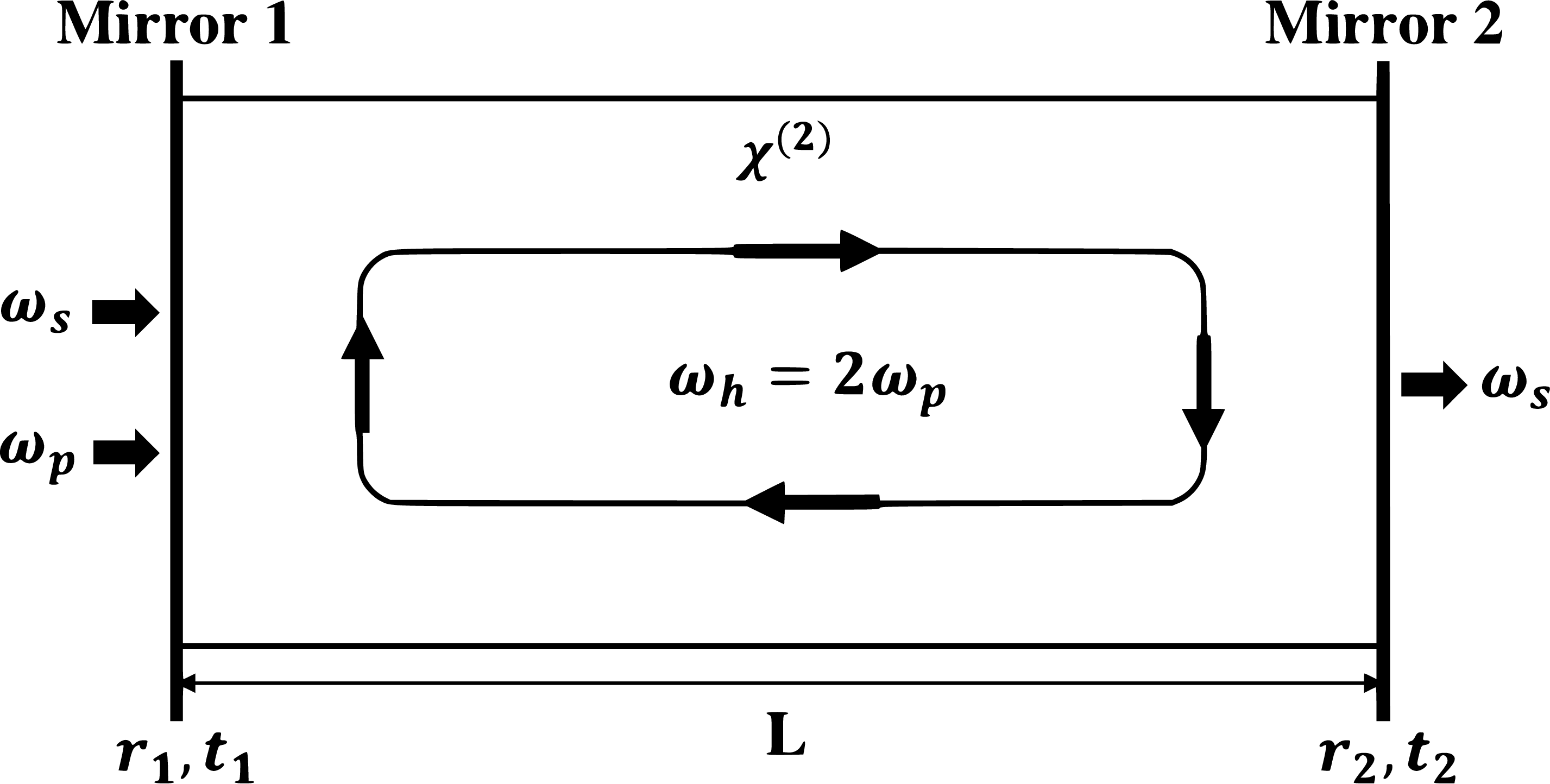}
\caption{A schematic diagram for the optical transistor 
operating with the waves of the single frequency 
in the nonlinear resonant structure.}
\label{fig:epsart-SHG/SHG-scheme} 
\end{figure}

For the physical implementation of the optical transistor 
operating with waves of a single frequency 
in the nonlinear resonant structure,
the second-order nonlinear interactions of the two Type-I SHGs
 are proposed in the present study.
The two Type-I SHGs inherently involve SHG and iSHG
which are indispensable for  the  amplification and for the switching 
in the optical transistor.
%
%
SHG
 is a second-order nonlinear optical process 
in which a fundamental wave of identical frequency 
and polarization 
interacts within a nonlinear medium 
 to generate a second harmonic wave at twice the frequency
($\omega_{p} + \omega_{p} \rightarrow \omega_{h} = 2\omega_{p}$).
In this work, 
SHG is explicitly referred to as Type-I SHG 
to distinguish it from Type-II SHG.
iSHG
 is an inverse phenomenon of SHG. 
In this process, a fundamental wave ($\omega_{s}$) 
and its second harmonic wave ($\omega_{h}$) 
interact with a nonlinear material, 
and their mixing leads to amplification of the fundamental wave 
($\omega_{h} - \omega_{s} \rightarrow \omega_{s} = \omega_{h}  -\omega_{s}$).
%
%
The proposed  SHG/iSHG interactions are the cascaded processes of 
SHG $\omega_{p} + \omega_{p} \rightarrow \omega_{h} = 2\omega_{p}$  
and iSHG $\omega_{h} - \omega_{s} \rightarrow \omega_{s} = \omega_{h} - \omega_{s}$ 
in the second-order nonlinear optical phenomena.
In these phenomena, 
each of a pump wave and a signal wave is independently 
coupled to a single second harmonic wave 
without direct coupling  between the two of the pump and the signal wave. 
The second harmonic wave generated in the SHG interaction resonates 
inside the structure
and serves as a source  in iSHG.
In Fig.~\ref{fig:epsart-SHG/SHG-scheme},
a schematic diagram for the optical transistor operating with waves 
of a single frequency 
in the nonlinear resonant structure is shown. 
 A waveguide with a nonlinear susceptibility $\chi^{(2)}$
 and an interaction length $L$  is  placed between two mirrors  
with  reflection   and  transmission coefficients $r_{1}$, $t_{1}$, $r_{2}$, and
$t_{2}$ for the second harmonic wave. 
The pump wave and the signal wave are incident on mirror 1 and
the amplified signal wave is transmitted through mirror 2,
but the second harmonic wave  is singly resonant between the two mirrors 
in the structure. 
In the scheme, 
if the pump wave from the emitter is first incident solely on mirror 1 
and passes through the nonlinear waveguide,
the second harmonic wave is generated.  
The second harmonic wave resonates inside the structure 
composed of the mirrors with $r_{1} = r_{2} = 100\%$ \cite{Kotlyar2021} 
and  increases its power through the SHG interaction with the pump wave. 
Then  the power of the incident pump wave  can be adjusted
 so that the entire pump wave is 
depleted and converted into the second harmonic wave. 
The output power of the optical waves exiting the collector 
in this case is null
and can be defined to be zero, indicating a digital off state.
Subsequently, 
the signal wave can be prepared, which has the same frequency 
but orthogonal polarization compared with the pump wave.
If the signal wave from the base is sequentially incident on the mirror 1 
and passes through the nonlinear waveguide,
the signal wave interacts with the resonant second harmonic wave 
so as to generate an idler wave by way of the iSHG interaction.
Since the pump wave frequency is equal to the signal wave frequency,
the frequencies of the pump wave, the signal wave, and the idler wave are identical. 
The idler wave emerging from the collector in this case 
is the amplified signal wave whose output power is not zero 
and can be defined to be one, indicating a digital on state.
It is assumed in the meantime that the two mirrors in the resonant structure 
reflect the  second harmonic wave exclusively, 
but transmit all other waves including the pump and the signal wave.  
 In this way, the base wave controls the emitter wave so that the collector wave 
 is manipulated for amplification and for switching.
In a nonlinear resonant structure, 
both a resonant wave and  single pass waves can mix and give rise 
to nonlinear interactions.
 If the number of resonant cycles, $n$, of a resonant wave is introduced 
 into the nonlinear equations describing interactions among single pass waves, 
then the single pass equations with the wave amplitudes indexed by $n$ 
can in general describe the nonlinear interactions
among a resonant wave and single pass waves.
Thus, the interaction among a resonant second harmonic wave and 
other single pass waves can be  simplified to the interaction  
among a single pass second harmonic wave and   other single pass waves
at a general stage in the resonant cycle $n$ \cite{Kim2011}.
 %
 
%
On the basis of the SHG equations and the resonant structure 
formalism \cite{Armstrong1962, Kim2011}, 
%
%
the governing equations for the two Type-I SHGs, 
proposed to describe the cascaded SHG/iSHG interactions 
in the $n$-th resonant cycle, can be represented as
\begin{widetext}
\begin{subequations}
  \label{twoType-ISHGs}
\begin{eqnarray}
\partial_{z} E_{p,n}(z) &=& -\frac{\alpha_{p}}{2}E_{p,n}(z)
                     + i\frac{2\omega_{p}}{n_{p}cN_{p}} \kappa_{p}^{SHG} E_{p,n}^{*}(z) E_{h,n}(z) e^{i\Delta \beta^{SHG}_{p} z},
\label{twoType-ISHGs_{p,n}(z)}
\\
\partial_{z} E_{s,n}(z) &=& -\frac{\alpha_{s}}{2}E_{s,n}(z)
                     + i\frac{2\omega_{s}}{n_{s}cN_{s}} \kappa_{s}^{SHG} E_{s,n}^{*}(z) E_{h,n}(z) e^{i\Delta \beta^{SHG}_{s} z},
\label{twoType-ISHGs_{s,n}(z)}
\\
\partial_{z} E_{h,n}(z) &=& -\frac{\alpha_{h}}{2}E_{h,n}(z)
                     + i\frac{2\omega_{h}}{n_{h}cN_{h}} \frac{\kappa_{hp}^{SHG}}{2} E_{p,n}^{2}(z) e^{-i\Delta \beta^{SHG}_{p} z}
   \nonumber \\ 
 & &  \hspace{20.0mm}
                     + \hspace{0.8mm}
                      i\frac{2\omega_{h}}{n_{h}cN_{h}} \frac{\kappa_{hs}^{SHG}}{2} E_{s,n}^{2}(z) e^{-i\Delta \beta^{SHG}_{s} z},
\label{twoType-ISHGs_{h,n}(z)}
\end{eqnarray}  
\end{subequations}
\end{widetext}
under the slowly varying envelope approximation.
Here $E_{i,n}(z)$, $\kappa_{i}^{SHG}$, $\alpha_{i}$, 
$\omega_{i}$, $N_{i}$, and $n_{i}$ (i=p, s, h, hp, or hs) denote 
a complex amplitude of an electric field defined as 
$\vec{E}_{i,n}(\vec{x},t) 
= \vec{E}_{i,n}(x,y) E_{i,n}(z) e^{i(\beta_{i} z - \omega_{i} t)} + c.c.$ 
from a real classical electric field, 
a nonlinear coupling constant defined as  
 $\kappa_{p}^{SHG} = \int dxdy {\vec{E}_{p}}^{*}(x,y) 
 \frac{\chi^{(2)}}{2} \vec{E}_{h}(x,y) {\vec{E}_{p}}^{*}(x,y)$ etc., 
a propagation loss constant, 
an angular frequency,
a normalization constant of integration defined as 
$N_{i}=\int dxdy|\vec{E}_{i,n}(x,y)|^{2} $, 
and  an effective index of refraction                 
corresponding to the pump wave, the signal wave, 
and the second harmonic wave, respectively. 
The phase factors 
$\Delta \beta^{SHG}_{p} = \beta_{h} - 2\beta_{p} - \beta_{\Lambda}$
 and   
$\Delta \beta^{SHG}_{s}=\beta_{h}-2\beta_{s}-\beta_{\Lambda}$        
are the SHG phase mismatches among the propagation constants
$\beta_{i}=\frac{\omega_{i}}{c}n_{i}(\omega_{i},\hat{e}_{i})$ 
and $\beta_{\Lambda}=\frac{2\pi N}{\Lambda}$.
Here $\hat{e}_{i}$,  $\Lambda$, $N$, and $c$ denote 
the polarization unit vector,
the period of quasi phase matching, 
the positive integer of a phase matching order, 
and the  speed of light in vacuum, respectively.
The propagation modes in the waveguide  
are determined from the decoupled wave equation
 $(\partial_{x}^{2} + \partial_{y}^{2} + \omega_{i}^{2} \mu_{0} \epsilon 
 - \beta_{i}^{2}) \vec{E}_{i,n}(x,y) = 0$, implying that 
 the solution $\vec{E}_{i,n}(x,y)$ is independent of the number $n$ in the function.
The nonlinear equations in Eqs.~(\ref{twoType-ISHGs}) are symmetric
 with respect to $ p \leftrightarrow s$ interchange 
and possess the two Type-I SHGs that can potentially 
contain both SHG and iSHG in themselves. 
%

%
\subsection{\label{OTR-1-Theoretical-analysis} Theoretical  analysis}
For a theoretical study of the nonlinear equations in Eqs.~(\ref{twoType-ISHGs}),   
the phase matching condition $\Delta \beta^{SHG}_{p}=\Delta \beta^{SHG}_{s}=0$ is
implicitly imposed on the nonlinear optical processes to be
discussed below. 
In particular, in the two Type-I SHGs, 
the phase matching  requires a waveguide condition
\begin{equation}
n_{p}(\omega_{p}, \hat{e}_{p}) = n_{s}(\omega_{s}, \hat{e}_{s})
\label{n_{eff}-omega_{p}-omega_{s}}
\end{equation}
for the two effective indices of refraction  
of the pump wave and the signal wave that possess 
the same frequency $\omega_{p} = \omega_{s}$ but orthogonal polarizations 
$\hat{e}_{p} \cdot \hat{e}_{s} = \delta_{p,s}$.
%
%
%
While uniaxial crystals such as LiNbO$_{3}$ intrinsically
possess identical bulk refractive indices along two orthogonal axes, 
achieving the condition 
in Eq.~(\ref{n_{eff}-omega_{p}-omega_{s}}) 
may constitute the most demanding challenge
in realizing a waveguide cavity with very low loss
for the practical implementation of the present optical transistor  
based on the nonlinear resonant structure.
%
%

%
  In the lossless limit $\alpha_{i} = 0$ for analytical convenience, 
the complex equations in Eqs.~(\ref{twoType-ISHGs}) can be written again 
in terms of 
real equations to be solved for exact theoretical  solutions. 
To this end, 
the complex amplitudes of the electric fields can be first represented as
\begin{subequations}
  \label{parameteru_{psh}}
\begin{eqnarray}
E_{p,n}(z) &=& \sqrt{\frac{\omega_{p}}{n_{p}cN_{p}}} u_{p,n}(z) e^{i\phi_{p,n} (z)},
\label{parameteru_{p,n}(z)}
\\
E_{s,n}(z) &=& \sqrt{\frac{\omega_{s}}{n_{s}cN_{s}}} u_{s,n}(z) e^{i\phi_{s,n} (z)},
\label{parameteru_{s,n}(z)}
\\
E_{h,n}(z) &=& \sqrt{\frac{\omega_{h}}{2n_{h}cN_{h}}} u_{h,n}(z) e^{i\phi_{h,n} (z)},
\label{parameteru_{h,n}(z)}
\end{eqnarray}
\end{subequations}
 and the nonlinear coupling constants can be replaced by 
\begin{subequations}
  \label{SHG-iSHG-K_{p}-K_{s}}
 \begin{eqnarray}
K_{p} &=& 2 \kappa_{p} 
\frac{\omega_{p}}{n_{p}cN_{p}}
\sqrt{\frac{\omega_{h}}{2n_{h}cN_{h}}},
\label{SHG-iSHG-K_{p}}
\\
K_{s} &=& 2 \kappa_{s} 
\frac{\omega_{s}}{n_{s}cN_{s}}
\sqrt{\frac{\omega_{h}}{2n_{h}cN_{h}}}.
\label{SHG-iSHG-K_{s}}
\end{eqnarray}
\end{subequations}
For the expressions in Eqs.~(\ref{SHG-iSHG-K_{p}-K_{s}}), 
the nonlinear coupling constants are defined as
$\kappa_{p}^{SHG} = \kappa_{p} e^{i\delta_{p}}$,  
$\kappa_{s}^{SHG} = \kappa_{s} e^{i\delta_{s}}$ 
and Kleinman symmetry \cite{Kleinmann1962} is  used for  the relations
$\kappa_{p}^{SHG}={\kappa_{hp}^{SHG}}^{*}$,        
$\kappa_{s}^{SHG}={\kappa_{hs}^{SHG}}^{*}$.         
Also, two parameters for the nonlinear phase difference
can be defined as 
\begin{subequations}
  \label{def-theta_{p,n}-theta_{s,n}}
\begin{eqnarray}
\theta_{p,n}(z) &=& \phi_{h,n}(z) - 2\phi_{p,n}(z) + \delta_{p},
\label{def-theta_{p,n}}
\\
\theta_{s,n}(z) &=& \phi_{h,n}(z) - 2\phi_{s,n}(z) + \delta_{s},
\label{def-theta_{s,n}}
\end{eqnarray}
\end{subequations}
to handle the phases of the complex electric fields 
and the nonlinear coupling constants.
Taking all of these into consideration, 
the nonlinear equations of the two Type-I SHGs 
in Eqs.~(\ref{twoType-ISHGs}) can be expressed as
\begin{subequations}
  \label{RealtwoType-ISHGs}
\begin{eqnarray}
\partial_{z} u_{p,n}(z) &=& -K_{p} u_{p,n}(z) u_{h,n}(z) \sin\theta_{p,n}(z),
\label{u_{p,n}}
\\
\partial_{z} u_{s,n}(z) &=& -K_{s} u_{s,n}(z) u_{h,n}(z) \sin\theta_{s,n}(z),
\label{u_{s,n}}
\\
\partial_{z} u_{h,n}(z) &=& K_{p} u_{p,n}^{2}(z) \sin\theta_{p,n}(z)  \nonumber   \\
                        &+& K_{s} u_{s,n}^{2}(z) \sin\theta_{s,n}(z),  
\label{u_{h,n}} 
\\
\partial_{z} \theta_{p,n}(z) &=& \frac{\Gamma_{n}(z)}{u_{h,n}^{2}(z)} + 2\frac{\cos\theta_{p,n}(z)}{\sin\theta_{p,n}(z)} \partial_{z} \ln u_{p,n}(z),
\label{theta_{p,n}}
\\
\partial_{z} \theta_{s,n}(z) &=& \frac{\Gamma_{n}(z)}{u_{h,n}^{2}(z)} + 2\frac{\cos\theta_{s,n}(z)}{\sin\theta_{s,n}(z)} \partial_{z} \ln u_{s,n}(z).
\label{theta_{s,n}}
\end{eqnarray}
\end{subequations}
For the sake of notational simplicity 
in Eqs.~(\ref{theta_{p,n}})--(\ref{theta_{s,n}}) 
and for later discussion, 
an additional  definition of a parameter $\Gamma_{n}(z)$ is introduced as 
\begin{eqnarray}
\Gamma_{n}(z) &=& K_{p} u_{p,n}^{2}(z) u_{h,n}(z) \cos\theta_{p,n}(z)  \nonumber   \\
              &+& K_{s} u_{s,n}^{2}(z) u_{h,n}(z) \cos\theta_{s,n}(z).
 \label{Gamma_{n}}
\end{eqnarray}
The real equations in Eqs.~(\ref{RealtwoType-ISHGs}) are the complete equivalence 
of the complex equations in Eqs.~(\ref{twoType-ISHGs}). 
The real equations 
 show that the amplitudes of the pump wave and the signal wave 
still remain decoupled from each other but coupled to the second harmonic wave, 
while the parameters of the nonlinear phase difference 
corresponding to the two waves are coupled 
with each other in the presence of $\Gamma_{n}(z)$.
Then, from  Eqs.~(\ref{RealtwoType-ISHGs}), 
conserved quantities 
for the two Type-I SHGs in the general $n$-th resonant cycle
can be derived for further analysis.
For the first constant of integration, 
the Manley--Rowe relation \cite{ManleyRowe1959} 
can be derived, leading to
\begin{equation} 
 \sum_{i}^{p,s,h} u_{i,n}^{2}(z) =  \sum_{i}^{p,s,h} u_{i,n}^{2}(0)
\label{Manley-Rowe-relation}
\end{equation}
from Eqs.~(\ref{u_{p,n}})--(\ref{u_{h,n}}). 
The relation in Eq.~(\ref{Manley-Rowe-relation}) 
is the conservation of total power flow in the waveguide 
with the power expression 
$P_{i,n}(z)=2\epsilon_{0} n_{i} c N_{i} |E_{i,n}(z)|^{2}$.
 As for the second constant of integration, 
 the parameter $ \Gamma_{n}(z)$ is 
an intrinsically conserved quantity, given by
\begin{eqnarray}
\partial_{z} \Gamma_{n}(z) &=& 0,
 \label{d-Gamma_{n}}
\end{eqnarray}
which can be derived from Eqs.~(\ref{theta_{p,n}})--(\ref{theta_{s,n}}). 
For the conserved quantity $\Gamma_{n}(z)$, 
  if $K_{p} \rightarrow 0$ or $K_{s} \rightarrow 0$, 
$\Gamma_{n}(z)$ reduces to the well known constant of integration  
for a single Type-I SHG in single pass propagation.
An additional third constant of integration can be derived 
 from Eqs.~(\ref{u_{p,n}})--(\ref{u_{s,n}}) as
\begin{eqnarray}
\frac{\partial_{z} \ln u_{p,n}(z)}{\partial_{z} \ln u_{s,n}(z)} 
&=& \frac{K_{p}}{K_{s}} \cdot \frac{\sin\theta_{p,n}(z)}{\sin\theta_{s,n}(z)}. 
 \label{third constant of integration}
\end{eqnarray}
The constancy of the relation in Eq.~(\ref{third constant of integration})
will be discussed in the next section. 
%

%
\subsection{\label{OTR-1-cSHG/iSHG-Solutions} Solutions of cascaded SHG/iSHG}
%

\begin{figure}
\includegraphics[width=8.0cm, angle=0]{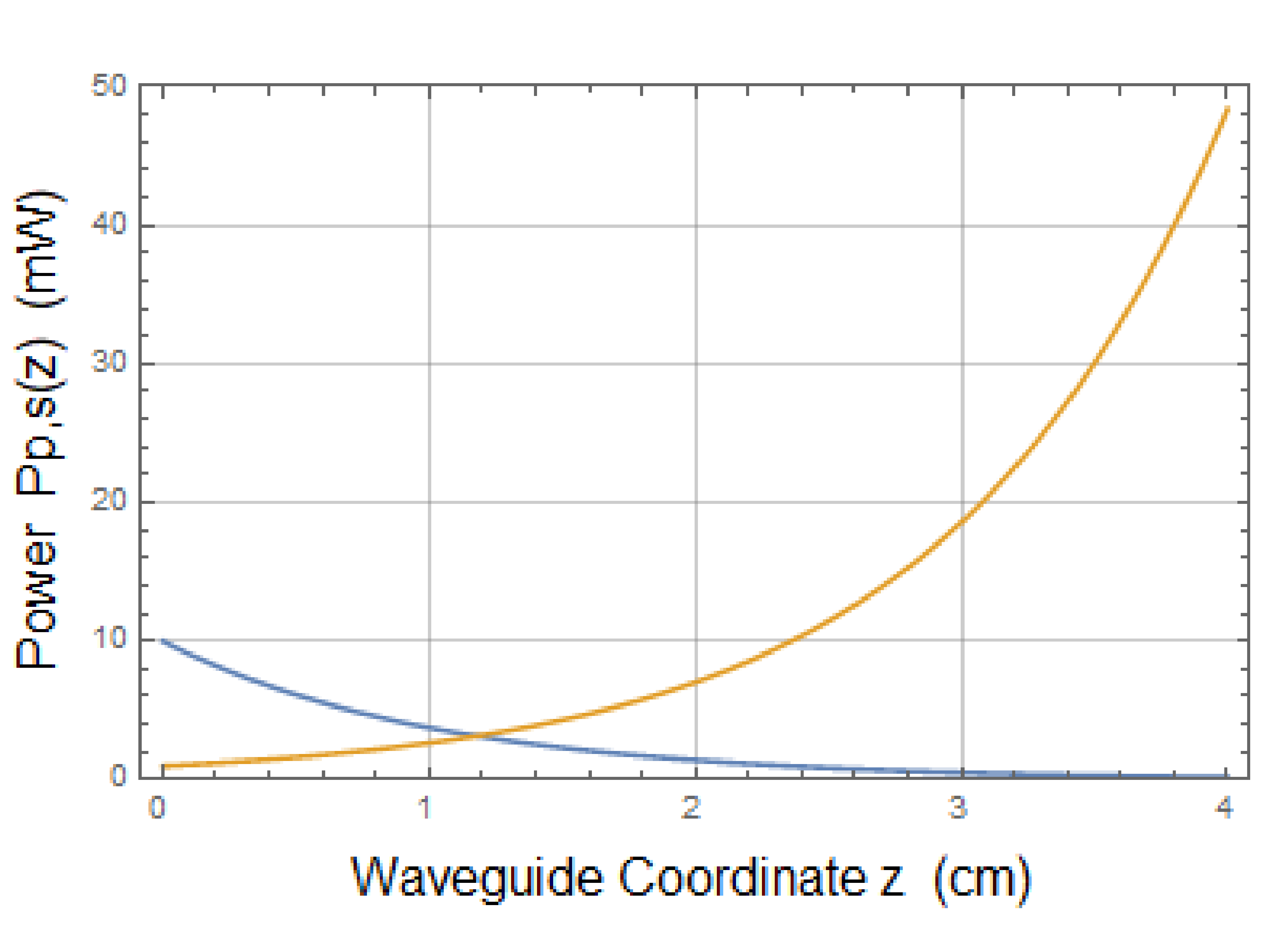}
\caption{The blue curve shows the deamplification of the pump wave due to SHG, 
while the yellow curve shows the amplification of the signal wave due to iSHG.}
\label{fig:epsart-SHG/iSHG-plotting}
\end{figure}

%
The nonlinear equations in Eqs.~(\ref{RealtwoType-ISHGs}) are 
essentially combined forms of two SHG equations for the pump wave 
and the signal wave 
in the absence of any direct coupling between the amplitudes of the two waves.
The equations involve  five unknown parameters--three amplitudes and two
nonlinear phase differences to be solved.
Nevertheless, 
the second constant of integration in  Eq.~(\ref{d-Gamma_{n}})  
provides the solutions for the two parameters of the nonlinear phase differences 
in Eqs.~(\ref{theta_{p,n}})--(\ref{theta_{s,n}}),  
   which  in turn can be used to solve the three amplitude equations 
in Eqs.~(\ref{u_{p,n}})--(\ref{u_{h,n}}).
If the pump wave is initially incident alone on the emitter and passes 
through the nonlinear waveguide, 
the initial condition for the second harmonic wave is
\begin{eqnarray}
 u_{h,0}(0) = 0
\label{u_{h,0}(0) = 0}
\end{eqnarray}
in the beginning of the SHG interaction and resonance.
This condition implies that  $\Gamma_{0}(z) = 0$ 
and then $\cos\theta_{p,0}(z) = 0$.
 Since the backward propagation  of the second harmonic wave 
 from $z=L$ to $z=0$
 does not change the nonlinear phases, it follows that
 $\cos\theta_{p,1}(0) = 0$, so that $\Gamma_{1}(z) = 0$ 
 and hence $\cos\theta_{p,1}(z) = 0$.
 This process can be applied to repeat during  the resonance, 
 increasing the number of resonant cycles in consecutive order, 
 to finally obtain
 $\cos\theta_{p,n}(z) = 0$ in the $n$-th cycle.
 In addition, the second harmonic wave is  assumed to be resonating 
 under the resonant feedback \cite{Ashkin1966, Fujimura1996} 
 and  the  backward propagation condition \cite{Kim2011}  
\begin{equation}
E_{h,n+1}(0)=r_{1}r_{2}e^{-\frac{\alpha_{h}}{2}L}e^{2i\beta_{h}L} E_{h,n}(L)
 \label{resonant-feedback-E_{h,n+1}(0)}
\end{equation}
where the resonant wave itself participates in the nonlinear interactions 
in each feedback.  
 If the pump wave is followed by the signal wave, 
 the nonlinear phase of the second harmonic wave is still preserved
 in the presence of the signal wave as well.
Even if the pump wave and the signal wave are incident at the same time, 
 the process is identical to the case as explained just above.
Therefore, the absence of the initial input for the second harmonic wave 
 in Eq.~(\ref{u_{h,0}(0) = 0})  leads to
\begin{eqnarray}
 \Gamma_{n}(z) = 0
\label{Gamma_{n}(z) = 0}
\end{eqnarray}
and thus $\cos\theta_{p,n}(z) = \cos\theta_{s,n}(z) = 0$,
because
 the condition   in Eq.~(\ref{Gamma_{n}(z) = 0})   is  maintained 
 for any arbitrary values of  $K_{i}$ and $u_{i,n}(z) ~ (i=p, s, h)$.
Consequently, 
the second constant of integration  provides
\begin{eqnarray}
 \theta_{p,n}(z) = \pm  \frac{\pi}{2}, ~ \theta_{s,n}(z) =  \pm  \frac{\pi}{2}
\label{theta_{p,n}(z)-theta_{s,n}(z)-solutions}
\end{eqnarray}
which are also the solutions of  Eqs.~(\ref{theta_{p,n}})--(\ref{theta_{s,n}}).
The nonlinear phase differences are constant and  conserved  
without the loss of generality in the resonance. 
Then the first constant of integration 
in Eq.~(\ref{Manley-Rowe-relation}) 
and the third constant of integration 
in Eq.~(\ref{third constant of integration}) 
can be used to obtain the solution for the remaining one amplitude parameter.
 Now, the exact theoretical solutions of the nonlinear equations 
 in Eqs.~(\ref{RealtwoType-ISHGs})
 can be obtained  in the following two cases. 
 In the first  scheme, if $\kappa_{p} = -\kappa_{s}$ 
with $\sin\theta_{p,n}(0) = \sin\theta_{s,n}(0)$, 
the third constant of integration in Eq.~(\ref{third constant of integration}) 
reduces to
\begin{eqnarray}
\frac{\partial_{z} \ln u_{p,n}(z)}{\partial_{z} \ln u_{s,n}(z)} = -1 
 \label{third-constant-of-integration-1}
\end{eqnarray}
and gives a relation between the amplitudes 
 of the pump wave and the  signal wave as
\begin{eqnarray}
u_{s,n}(z) = u_{s,n}(0) \frac{u_{p,n}(0)}{u_{p,n}(z)}. 
\label{amplitude-product-1}
\end{eqnarray}
The relation in Eq.~(\ref{amplitude-product-1}) shows that the amplitude 
of the signal wave is inversely proportional 
to the amplitude of the pump wave. 
Therefore, it  explicitly indicates that the signal wave can be amplified 
if the pump wave is deamplified. 
  Along with the relations from the three constants of integration,
 after some algebra, analytical solutions of the cascaded SHG/iSHG 
 can be obtained from  Eqs.~(\ref{u_{p,n}})--(\ref{u_{h,n}})  as 
\begin{widetext}
\begin{subequations}
  \label{solutions-of-SHG/iSHG}
\begin{eqnarray}
u_{p,n}(z) &=& \frac{U_{n}(0)}{\sqrt{2}\sqrt{1+k_{n}^{2}}} 
[\textnormal{dn} \{b (z+z_{n}), k_{n}\} 
+ k_{n} \textnormal{cn} \{b (z+z_{n}), k_{n}\}],
\label{solutions-of-SHG/iSHG-u_{p,n}}
\\
u_{s,n}(z) &=& \frac{U_{n}(0)}{\sqrt{2}\sqrt{1+k_{n}^{2}}} 
[\textnormal{dn} \{b (z+z_{n}), k_{n}\} 
- k_{n} \textnormal{cn} \{b (z+z_{n}), k_{n}\}],
\label{solutions-of-SHG/iSHG-u_{s,n}}
\\
u_{h,n}(z) &=& \frac{\sqrt{2} k_{n} U_{n}(0)}{\sqrt{1+k_{n}^{2}}} 
\textnormal{sn} [b (z+z_{n}), k_{n}].
\label{solutions-of-SHG/iSHG-u_{h,n}}
\end{eqnarray}
\end{subequations}
\end{widetext}
Here, the expressions sn(z,k), cn(z,k), dn(z,k), 
and sc(z,k) in Eq.~(\ref{z_{n}-SHG/iSHG}) below 
 stand for Jacobi elliptic functions \cite{Jacobi1829, Schwalm15}.
For the simplification of the solutions, 
four constants $U_{n}(0)$, $k_{n}$, $b$, and $z_{n}$ are introduced 
in the expressions.
The constant $U_{n}(0)$, which can be written as 
\begin{eqnarray}
U_{n}(0) &=& \sqrt{u_{p,n}^{2}(0) + u_{s,n}^{2}(0) + u_{h,n}^{2}(0)},
\label{U_{n}(0)-SHG/iSHG}
\end{eqnarray}
plays the role of the amplitude in each Jacobi elliptic function 
in the solutions.  
The modulus $k_{n}$, which can be defined as 
\begin{eqnarray}
k_{n}^{2} = \frac{U_{n}^{2}(0) - 2u_{p,n}(0) u_{s,n}(0)}
                  {U_{n}^{2}(0) + 2u_{p,n}(0) u_{s,n}(0)} 
\label{k_{n}-SHG/iSHG} 
\end{eqnarray}
with $0 \leq k_{n} \leq 1$,
determines the characteristics of the Jacobi elliptic functions and
 the parameter $b$ can be represented as
\begin{eqnarray}
b  = \frac{\sqrt{2} K_{p} \sin\theta_{p,n}(0) U_{n}(0)}
                  {\sqrt{1+k_{n}^{2}}}. 
\label{b-SHG/iSHG}  
\end{eqnarray}
The spatial displacement $z_{n}$ in the presence of $u_{h,n}(0)$ 
  in the $n$-th resonant cycle  can be given by
\begin{eqnarray}
z_{n} = \frac{1}{b} \textnormal{sc}^{-1} 
[\frac{u_{h,n}(0)}{u_{p,n}(0) - u_{s,n}(0)}, k_{n} ]  
\label{z_{n}-SHG/iSHG}
\end{eqnarray}
 in the present scheme.
With all of the parameters, 
the expressions  in Eq.~(\ref{theta_{p,n}(z)-theta_{s,n}(z)-solutions}) 
and Eqs.~(\ref{solutions-of-SHG/iSHG}) 
are the exact solutions for the cascaded SHG/iSHG with 
$K_{p} = -K_{s}$ in Eqs.~(\ref{RealtwoType-ISHGs}). 
From the point of view concerned with the behaviors 
of the Jacobi elliptic functions,
the solutions suggest that, 
in the range $0 \leq z \leq L$ of the nonlinear resonant structure,
the pump wave should be deamplified due to 
the summation of the functions dn(z,k) and cn(z,k), 
while the signal wave should be amplified due to 
the difference between the functions dn(z,k) and cn(z,k).
  In the limit as $k_{n} \rightarrow 0$ in the solutions,  a condition 
\begin{eqnarray} 
u_{p,n}(0) = u_{s,n}(0), ~ u_{h,n}(0) = 0
\label{SHG/iSHG-transparency-conditions}
\end{eqnarray}
is  derived from Eqs.~(\ref{U_{n}(0)-SHG/iSHG})--(\ref{k_{n}-SHG/iSHG}), 
and in this case, the condition leads to the solutions
\begin{eqnarray} 
u_{i,n}(z) &=& u_{i,n}(0), ~ (i=p,s,h)
\label{SHG/iSHG-transparent-solutions}
\end{eqnarray}
 from  Eqs.~(\ref{solutions-of-SHG/iSHG}).
 The solutions in Eq.~(\ref{SHG/iSHG-transparent-solutions}) show  
 a phenomenon of nonlinear transparency
 that illustrates the transparent propagation of the  pump wave and the signal 
 wave without the nonlinear interactions in the  waveguide.
The nonlinear transparency is due to the balance and the equilibrium 
between SHG and iSHG 
as can be proved from Eqs.~(\ref{u_{p,n}})--(\ref{u_{h,n}}). 
The SHG/iSHG transparency can be confirmed, as well,
 by the numerical calculations  of the complex equations 
 in Eqs.~(\ref{twoType-ISHGs}) 
 with the initial conditions of the three waves,
  for example $P_{p,n}(0) = P_{s,n}(0) = 10$ mW, $P_{h,n}(0) =0$,
  as required  by Eq.~(\ref{SHG/iSHG-transparency-conditions}).
It  can be used for a verification 
of the waveguide condition in Eq.~(\ref{n_{eff}-omega_{p}-omega_{s}}) 
with or without a resonant structure under the present consideration.   
In the limit as $k_{n} \rightarrow 1$,  
  Eqs.~(\ref{solutions-of-SHG/iSHG}) reduce to the solutions of 
a single Type-I SHG driven by the pump wave with $u_{s,n}(z) = 0$ 
which implies a condition $u_{s,n}(0) = 0$  
imposed by Eq.~(\ref{k_{n}-SHG/iSHG}).
Numerical solutions in the scheme of $\kappa_{p} = -\kappa_{s}$
can be obtained from the direct calculations of the complex equations 
in Eqs.~(\ref{twoType-ISHGs}) 
with  $\kappa_{p}^{SHG} = -\kappa_{s}^{SHG}$ 
under  the condition in Eq.~(\ref{resonant-feedback-E_{h,n+1}(0)}).
These types of solutions can include behaviors of solutions 
of the nonlinear equations 
where the signs of the nonlinear coupling constants are opposite 
such as $\kappa_{p}^{SHG} \cdot  \kappa_{s}^{SHG} < 0$. 
For the convenience of numerical calculations, 
it is assumed that  
$\kappa_{p}^{SHG}/N_{p} = \kappa_{hp}^{SHG}/N_{h} = d_{21} \cdot \eta_{eff}$,
$\kappa_{s}^{SHG}/N_{s} = \kappa_{hs}^{SHG}/N_{h} = d_{22} \cdot \eta_{eff}$,
and 
the signal wave sequentially follows the pump wave 
that is  first incident alone to generate the resonant second harmonic wave.
Here
$d_{21}$ and $d_{22}$ stand for the elements of the d-coefficient matrix
and 
$\eta_{eff}$ is the effective mode factor 
including the overlap integral of waves. 
Then the pump wave polarized along the X-axis 
and the signal wave polarized along the Y-axis
are coupled with the second harmonic wave polarized along the Y-axis, 
where
X, Y, and Z are the principal coordinate axes in the crystal frame. 
For the calculations of solutions and the plotting, physical parameters
$-d_{21} = d_{22} = 3$ pm/V,
$\eta_{eff} = 0.9$,
$\alpha_{h} = 3\alpha_{p} = 0.003$ dB/cm,
wavelengths $\lambda_{p} = \lambda_{s} = 1550$ nm for $\omega_{p} = \omega_{s}$, 
$n_{p} = n_{s} =  2.1544$, $n_{h} = 2.2325$, and
 $L = 4$ cm  are adopted. 
Together with the inputs 
$P_{p,n}(0) = 10$ mW, 
$P_{s,n}(0) = 1$ mW, 
$N_{i} = 1$ $\mu$m$^{2}$, and $n=100$, 
the power of the two waves $P_{p,s}(z)$ from the solutions 
is plotted 
as a function of waveguide coordinate $z$ in the laboratory frame 
 in Fig.~\ref{fig:epsart-SHG/iSHG-plotting},
including the factor  $\frac{2}{\pi}$ from quasi-phase matching.
Here, $N_{i} = 1$ $\mu$m$^{2}$ is assumed 
only for analytical convenience; 
the present formalism is not restricted to this cross-sectional area 
and applies equally to general optical waveguides with arbitrary 
$N_{i}$.
The notation $P_{p,s}(z)$ denotes the plotted power traces 
of the pump and signal, i.e., $P_{p,s}(z)\equiv\{P_{p,n} (z),\,P_{s,n} (z)\}$.
The deamplification of the pump wave due to SHG that 
can define the digital off state
and the amplification of the signal wave 
due to iSHG that can define the digital on state
are demonstrated in the figure. 
The numerical solutions of  Eqs.~(\ref{twoType-ISHGs}) 
constitute an independent confirmation of the behaviors 
of the analytical solutions 
in Eqs.~(\ref{solutions-of-SHG/iSHG}).
%

%
%
Regarding input power limits, the operational power limits 
are determined primarily 
by the properties of the reflection mirrors 
and the waveguide. 
The SiN semiconductor DBRs and the LiNbO$_{3}$ crystal waveguide 
are promising candidates as they exhibit high optical damage thresholds. 
Representative reported values for LiNbO$_{3}$ crystal waveguides 
include 26~MW/cm$^{2}$ \cite{Chen2007},
$2.36 \times 10^{5}$~W/cm$^{2}$ \cite{Tian2024},
and 300--500~MW/cm$^{2}$ in recent commercial products \cite{InradOptics2025}.
These reported thresholds are sufficient to sustain the resonant wave power 
calculated in this work, namely 
$P_{h,n}(L)/P_{p,n}(0) \simeq 69.1$ for $L=4$~cm in the cascaded SHG/iSHG scheme 
and $P_{h,n}(L)/P_{p,n}(0) \simeq 53.7$ for $L=1$~cm 
to be discussed later in the cascaded SHG/OPA scheme.
%

%
There is no widely accepted definition of efficiency for optical transistors yet. 
However, the power transfer ratio 
and the power amplification factor 
can serve as appropriate figures of merit to represent 
the efficiency of the proposed device.
%
%
Similar to the characteristics of an electronic transistor, 
 in the optical transistor of the present study
the power transfer ratio $\alpha_{TR}$ can be defined as 
\begin{eqnarray}
\alpha_{TR} = \frac{P_{s,n}(L)}{P_{p,n}(0)}
 \label{power-transfer-ratio}
\end{eqnarray}
to represent the ratio of the output power of the signal wave out of the collector
to the input power of the pump wave incident on the emitter. 
Likewise, the power amplification factor $\beta_{AF}$ can be defined as 
\begin{eqnarray}
\beta_{AF} =  \frac{P_{s,n}(L)}{P_{s,n}(0)}
 \label{power-amplification-factor}
\end{eqnarray}
to represent the ratio of the output power of the signal wave out of the collector
to the input power of the signal wave incident on the base. 
From the numerical results of the collector power 
$P_{s,n}(L) = 48.38$ mW,
the power transfer ratio $\alpha_{TR} = 4.838$  and  
the power amplification factor $\beta_{AF} = 48.38$
 can be obtained  in the present calculations. 
 %
 
%
Although the waveguide length $L=4$~cm, 
as indicated by the numerical analysis above and as shown
 in Fig.~\ref{fig:epsart-SHG/iSHG-plotting},
may appear long for practical applications, 
more compact form factors can be achieved through several approaches. 
Among these, artificial enhancement of $d_{21}$ and $d_{22}$ 
during the growth of LiNbO$_{3}$ 
constitutes an essential approach, as it directly strengthens 
the nonlinear interaction underlying the transistor operation. 
Alternatively, the use of materials possessing intrinsically 
larger nonlinear coefficients 
would also enable shorter devices. 
In addition, the use of a switching deadband enables a broader definition 
of the digital off state, thereby making shorter lengths feasible. 
%
%
Regarding the switching deadband, the deadband width 
$\Delta P = P_{\text{on}} - P_{\text{off}}$ 
can serve as a useful figure of merit. 
As one example, in the present device this reduces to 
$\Delta P = P_{p,n}(0) - P_{p,n}(0)/2 = P_{p,n}(0)/2$, 
which scales directly with the input pump power.
For instance, $\Delta P = 5~\mathrm{mW}$ when $P_{p,n}(0) = 10~\mathrm{mW}$ 
and $\Delta P = 0.5~\mathrm{mW}$ when $P_{p,n}(0) = 1~\mathrm{mW}$.
%
%
Therefore, 
if the off state is defined below $P_{\text{off}} = P_{p,n}(0)/2=5$~mW
for the sole input of the pump wave, 
practical operation can be realized with $L \geq 1$ cm,
%
because the output collector power is calculated to be
$P_{p,n}(L)$ = 8.1643, 4.9502, 1.3737, 0.391, 0.124, 0.0433 mW 
for $L$ = 0.5, 1.0, 2.0, 3.0, 4.0, 5.0 cm, respectively.
The input signal power can then be independently chosen 
so that the power transfer ratio 
$\alpha_{TR}=P_{s,n}(L)/P_{p,n}(0) \geq 2$, 
thereby satisfying the fan-out criterion. 
%
Design modifications to the waveguide that increase the effective mode factor 
as $\eta_{eff}\rightarrow 1$ will also help shorten the waveguide 
toward a compact form.
%
%
These considerations indicate that the proposed structure, 
though initially demonstrated with $L=4$~cm, 
can be realistically adapted to more compact dimensions 
through parameter optimizations and material choices.
%
  
%
\subsection{\label{OTR-1-cSHG/SHG-Solutions}  Solutions of cascaded SHG/SHG}
In the second scheme, if $\kappa_{p} = \kappa_{s}$ 
with $\sin\theta_{p,n}(0) = \sin\theta_{s,n}(0)$,   
the third constant of integration 
in Eq.~(\ref{third constant of integration}) reduces to
\begin{eqnarray}
\frac{\partial_{z} \ln u_{p,n}(z)}{\partial_{z} \ln u_{s,n}(z)} = 1 
 \label{third-constant-of-integration+1}
\end{eqnarray}
and gives a relation between the amplitudes
 of the pump wave and the  signal wave expressed as
\begin{eqnarray}
u_{s,n}(z) = u_{s,n}(0) \frac{u_{p,n}(z)}{u_{p,n}(0)}. 
\label{amplitude-product+1}
\end{eqnarray}
The relation in Eq.~(\ref{amplitude-product+1}) shows 
that the signal wave amplitude is linearly proportional 
to the pump wave amplitude.
Therefore, it explicitly indicates that the signal wave 
can be deamplified if the pump wave is deamplified. 
After some algebra similar to the first scheme, 
 analytical solutions of the cascaded SHG/SHG can be obtained from 
 Eqs.~(\ref{u_{p,n}})--(\ref{u_{h,n}})  as 
%
\begin{subequations}
  \label{solutions-of-SHG/SHG}
\begin{eqnarray}
u_{p,n}(z) &=& 
 \frac{u_{p,n}(0) U_{n}(0) \textnormal{cn}[b (z+z_{n}), k_{n}=1]} 
 {\sqrt{u_{p,n}^{2}(0) + u_{s,n}^{2}(0)}},
\label{solutions-of-SHG/SHG-u_{p,n}}
\\
u_{s,n}(z) &=& 
 \frac{u_{s,n}(0) U_{n}(0)\textnormal{cn}[b (z+z_{n}), k_{n}=1]} 
 {\sqrt{u_{p,n}^{2}(0) + u_{s,n}^{2}(0)}} ,
\label{solutions-of-SHG/SHG-u_{s,n}}
\\
u_{h,n}(z) &=& U_{n}(0) \textnormal{sn}[b (z+z_{n}), k_{n}=1].
\label{solutions-of-SHG/SHG-u_{h,n}}
\end{eqnarray}
\end{subequations}
%
%
Here, 
two constants $U_{n}(0)$ and $b$ are adopted from the expressions
in Eq.~(\ref{U_{n}(0)-SHG/iSHG}) and Eq.~(\ref{b-SHG/iSHG}), respectively.
The spatial displacement $z_{n}$ can be given as
\begin{eqnarray}
z_{n} = \frac{1} {b} \sinh^{-1} [\frac{u_{h,n}(0)} 
{\sqrt{u_{p,n}^{2}(0) + u_{s,n}^{2}(0)}}]
  \label{z_{n}-SHG/SHG}
\end{eqnarray}
 in the present scheme.
The expressions in Eq.~(\ref{theta_{p,n}(z)-theta_{s,n}(z)-solutions}) 
and Eqs.~(\ref{solutions-of-SHG/SHG}) 
are the exact solutions 
for the cascaded SHG/SHG with $K_{p} = K_{s}$  in Eqs.~(\ref{RealtwoType-ISHGs}).
The solutions in Eqs.~(\ref{solutions-of-SHG/SHG}) include
two SHG interactions for the simultaneous deamplifications 
of the pump wave and the signal wave
with  only the inherent  $k_{n}=1$.
In the limit $u_{p,n}(0) \rightarrow 0$ or $u_{s,n}(0) \rightarrow 0$, 
 Eqs.~(\ref{solutions-of-SHG/SHG}) reduce to the well known solutions 
 of a single Type-I SHG interaction. 
Numerical solutions in the scheme  of $\kappa_{p} = \kappa_{s}$
can be obtained from the direct calculations of the complex equations 
in Eqs.~(\ref{twoType-ISHGs}) 
with  $\kappa_{p}^{SHG} = \kappa_{s}^{SHG}$ and  the condition 
in Eq.~(\ref{resonant-feedback-E_{h,n+1}(0)}).
These types of solutions can include behaviors of solutions 
of the nonlinear equations 
where the signs of the nonlinear coupling constants are identical 
such as $\kappa_{p}^{SHG} \cdot  \kappa_{s}^{SHG} > 0$. 
Before the calculations,
let us first consider the relationship between the two  
$\kappa_{p} = - \kappa_{s}$ and $\kappa_{p} =  \kappa_{s}$ 
cases in the nonlinear  equations.
In the case of $\kappa_{p}^{SHG} = \pm \kappa_{s}^{SHG}$,  
the two sets of the complex  equations can be related by the amplitude 
transformation of the signal wave
\begin{equation}
  E_{s,n}(z) \longrightarrow  \pm i E_{s,n}(z) 
  \label{transformation-E_{s,n}(z)} 
\end{equation}
in Eqs.~(\ref{twoType-ISHGs}),
or equivalently, the corresponding sets of the real equations 
 with $K_{p} = \pm K_{s}$ 
can be related by the   nonlinear phase  transformation of the signal wave
\begin{equation}
 \phi_{s,n}(z) \longrightarrow \phi_{s,n}(z) \pm \frac{\pi}{2}  
  \label{transformation-phi_{s,n}(z)} 
\end{equation}
in Eqs.~(\ref{RealtwoType-ISHGs}).
This means that, 
in the two schemes with the nonlinear coupling constants mentioned above, 
the two sets of the nonlinear equations  for the cascaded SHG/iSHG 
and the cascaded SHG/SHG
can be transformed from one to the other
by means of the  transformation 
in Eq.~(\ref{transformation-E_{s,n}(z)}) or Eq.~(\ref{transformation-phi_{s,n}(z)}).
However, the implications of 
Eq.~(\ref{def-theta_{s,n}}),  
Eqs.~(\ref{RealtwoType-ISHGs}), and Eq.~(\ref{Gamma_{n}(z) = 0}) 
suggest that
the nonlinear phase shift of the signal wave
$\phi_{s,n}(0) \rightarrow \phi_{s,n}(0) \pm \frac{\pi}{2}$
at the initial input 
should lead to the same effect as  Eq.~(\ref{transformation-phi_{s,n}(z)})
and  consequently, the initial amplitude change   
$E_{s,n}(0) \rightarrow  \pm i E_{s,n}(0)$ in Eqs.~(\ref{twoType-ISHGs}) 
leads to the same effect  as  Eq.~(\ref{transformation-E_{s,n}(z)}).
Therefore,  the numerical solutions for the deamplification of the pump wave and
the amplification of the signal wave  can be obtained
from the cascaded SHG/SHG equations with $\kappa_{p}^{SHG} = \kappa_{s}^{SHG}$ 
and equivalently, 
 the numerical solutions for the simultaneous deamplifications 
 of the pump and the signal wave can be obtained
 from the cascaded SHG/iSHG equations with 
 $\kappa_{p}^{SHG} = -\kappa_{s}^{SHG}$ as well.
For the numerical solutions of Eqs.~(\ref{twoType-ISHGs}) 
with  $\kappa_{p}^{SHG} = \kappa_{s}^{SHG}$ to include the deamplification 
of the pump wave and the amplification of the signal wave,
all the parameters in the $\kappa_{p}^{SHG} = -\kappa_{s}^{SHG}$ calculations are
used  as they are, but  $d_{21} = d_{22} = 3$ pm/V 
and $P_{s,n}(0) = $ 1 mW with the phase-shifted  amplitude
$e^{\pm i\frac{\pi}{2}}  E_{s,n}(0)$ 
are adopted.
As discussed in Eq.~(\ref{transformation-E_{s,n}(z)}),
the numerical solutions in this   $\kappa_{p}^{SHG} = \kappa_{s}^{SHG}$ case
show that they are consistent with the numerical solutions 
 in the  $\kappa_{p}^{SHG} = -\kappa_{s}^{SHG}$ case
 and provide the same graph as in Fig.~\ref{fig:epsart-SHG/iSHG-plotting}. 
In the same manner as discussed above, the numerical solutions 
for simultaneous yet independent deamplifications of the pump wave and the signal wave 
can be demonstrated in Eqs.~(\ref{twoType-ISHGs})
 with $\kappa_{p}^{SHG} = -\kappa_{s}^{SHG}$ and the $\pm \frac{\pi}{2}$ 
 phase shift of the signal input amplitude, as well as
 with $\kappa_{p}^{SHG} = \kappa_{s}^{SHG}$ and the zero phase shift  
 of  the initial input amplitudes.
 The solutions in these two cases are also consistent with each other. 
In a typical second-order nonlinear material like LiNbO$_{3}$ 
to implement  the optical transistor as discussed in this section, 
the second-order nonlinear coefficient $d_{26} = 0$.
This means that there is no interaction caused by a direct coupling 
among the three waves participating in the Type-I SHG/iSHG interactions 
through $d_{21}$ and $d_{22}$
with the second harmonic wave of Y-polarization.
Since the second-order nonlinear coefficient $d_{16} \neq 0$  in general, 
however, 
the three waves can simultaneously participate in the Type-II SHG interaction 
through $d_{16}$
with the second harmonic wave of X-polarization.
%
%
%
In this context,
Type-II SHG
 is a second-order nonlinear optical process 
in which two fundamental waves of identical frequency ($\omega_{p} = \omega_{s}$) 
but orthogonal polarization ($\hat{e}_{p} \cdot \hat{e}_{s} = 0$) 
interact within a nonlinear material to generate a second harmonic wave 
at twice the frequency 
($\omega_{p} + \omega_{s} \rightarrow \omega_{h} = \omega_{p} + \omega_{s}$).
%
%
The generation of the second harmonic wave 
due to  $d_{16}$ reduces the amplification of the signal wave.
If the SHG interaction is carried out by way of the sole pump wave 
before the signal wave input, 
the power  of the second harmonic wave resonating with Y-polarization is much
stronger than the power  of the pump wave input 
such as $P_{h,n}(0) / P_{p,n}(0)$ $\simeq$ 66.5 
in the numerical calculations of Fig.~\ref{fig:epsart-SHG/iSHG-plotting}.
Thus 
the power  of the signal wave in the presence of 
$d_{16}(= d_{21} = -d_{22})$ relative to 
the power  of the signal wave in the absence of $d_{16}$ 
is not much reduced 
to  $P_{s,n}^{d_{16} = d_{21}}(L) /P_{s,n}^{d_{16} = 0}(L) = 98.7\% $ 
in the present calculations.
Hence even in the presence of $d_{16}$,
an overall strong amplification of the signal wave is anticipated 
because the amplification degree of the signal wave 
due to the Type-I SHG/iSHG interactions
is still greater than the deamplification degree of the signal wave 
due to the Type-II SHG interaction.
If possible in an artificial way, 
it would be best to make $d_{16} = 0$ during the growth process 
of a nonlinear material such as LiNbO$_{3}$ 
to avoid the Type-II SHG interaction.
%

%
\section{\label{OTR-2} An optical transistor  
\\  with waves of dual frequencies}

\subsection{\label{OTR-2-cSHG/OPA-theory} Theory of cascaded SHG/OPA}
%
%

\begin{figure}
\includegraphics[width=8.0cm, angle=0]{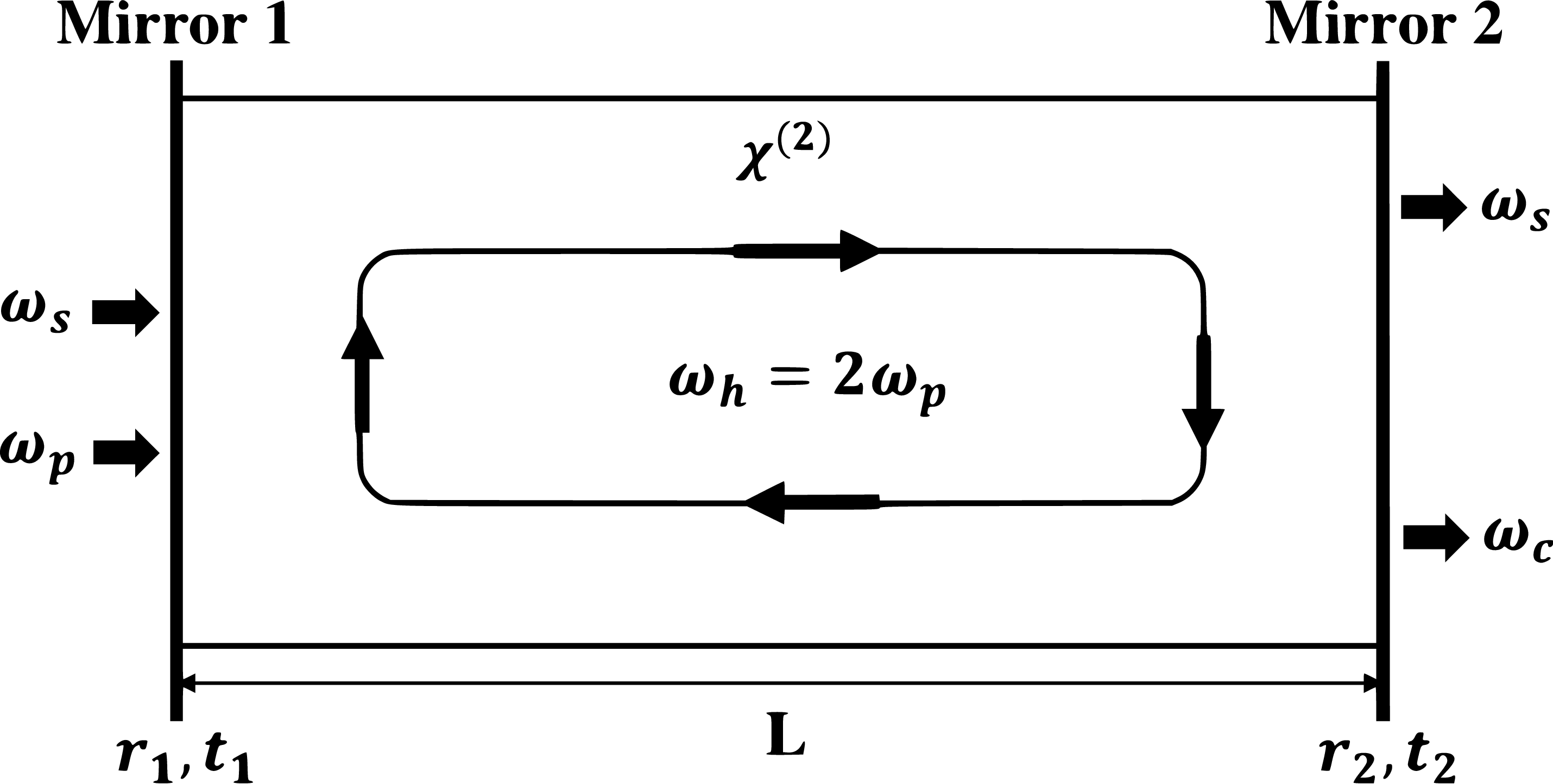}
\caption{\label{fig:epsart-SHG/OPA-scheme} 
A schematic diagram for the optical transistor operating with the waves 
of the dual frequencies in the nonlinear resonant structure.}
\end{figure}

%
In this section, a distinct type of optical transistor 
employing waves of dual frequencies through cascaded SHG and OPA processes 
is introduced.
For the physical implementation of the optical transistor 
operating with waves of  dual frequencies 
in the nonlinear resonant structure,
the second-order nonlinear interactions of 
 SHG and OPA are needed for amplification and switching.
%
%
In this context,
OPA is a second-order nonlinear optical process 
in which two waves of different frequencies ($\omega_{h}$ and $\omega_{s}$) 
interact within a nonlinear medium, 
resulting in amplification of the fundamental wave ($\omega_{s}$) 
and the generation of a new idler wave ($\omega_{c}$) at the difference frequency 
($\omega_{h} - \omega_{s} \rightarrow \omega_{c} = \omega_{h}  -\omega_{s}$). 
%
%
The cascaded SHG/OPA interactions are the processes of 
SHG $\omega_{p} + \omega_{p} \rightarrow \omega_{h} = 2\omega_{p}$
and OPA $\omega_{h} - \omega_{s} \rightarrow \omega_{c} = \omega_{h} - \omega_{s}$ 
in the second-order nonlinear phenomena. 
In Fig.~\ref{fig:epsart-SHG/OPA-scheme},
a schematic diagram for the optical transistor 
operating with the waves of the dual frequencies is illustrated. 
The nonlinear resonant structure in Fig.~\ref{fig:epsart-SHG/OPA-scheme} 
is the same as the one in Fig.~\ref{fig:epsart-SHG/SHG-scheme},
but in the present scheme, 
a converted idler wave is generated to come out of the collector
along with the signal wave.
 Since the pump wave frequency is different from  the signal wave frequency,  
$\omega_{p} \neq \omega_{s} $,
the frequencies of the pump, the signal, and the idler waves are distinguishable.
For the descriptions of the cascaded SHG/OPA interactions 
in the n-th resonant cycle,
the governing equations in the waveguide 
%
%
\cite{Armstrong1962, Gallo1997, Kim2011}
%
%
can be represented as
\begin{widetext}
\begin{subequations}
  \label{Cascaded-SHG/OPA} 
\begin{eqnarray}
\partial_{z} {E_{p,n}(z)} &=& -\frac{\alpha_{p}}{2} E_{p,n}(z)
                     + i\frac{2\omega_{p}}{n_{p}cN_{p}} \kappa_{p}^{SHG} E_{p,n}^{*}(z) E_{h,n}(z) e^{i\Delta \beta^{SHG} z},
\label{Cascaded-SHG/OPA-E_{p,n}}
\\
\partial_{z} {E_{h,n}(z)} &=& -\frac{\alpha_{h}}{2}E_{h,n}(z)
                     + i\frac{2\omega_{h}}{n_{h}cN_{h}} \frac{\kappa_{h}^{SHG}}{2} E_{p,n}^{2}(z) e^{-i\Delta \beta^{SHG} z} \nonumber
\\
  & &  \hspace{20mm}
                    + i\frac{2\omega_{h}}{n_{h}cN_{h}} \kappa_{h}^{OPA} E_{s,n}(z) E_{c,n}(z) e^{-i\Delta \beta^{OPA} z},
\label{Cascaded-SHG/OPA-E_{h,n}}
\\
\partial_{z} {E_{s,n}(z)} &=& -\frac{\alpha_{s}}{2} E_{s,n}(z)
                     + i\frac{2\omega_{s}}{n_{s}cN_{s}} \kappa_{s}^{OPA} E_{c,n}^{*}(z) E_{h,n}(z) e^{i\Delta \beta^{OPA} z},
\label{Cascaded-SHG/OPA-E_{s,n}}
\\
\partial_{z} {E_{c,n}(z)} &=& -\frac{\alpha_{c}}{2} E_{c,n}(z)
                     + i\frac{2\omega_{c}}{n_{c}cN_{c}} \kappa_{c}^{OPA} E_{s,n}^{*}(z) E_{h,n}(z) e^{i\Delta \beta^{OPA} z},
\label{Cascaded-SHG/OPA-E_{c,n}}
\end{eqnarray}
\end{subequations}
\end{widetext}
in the slowly varying envelope approximation.
  All the physical parameters are defined in the same way 
as  in Eqs.~(\ref{twoType-ISHGs}), 
  but   here the relevant fields are  the pump, 
  the second harmonic, the signal, and the converted idler wave. 
The phase factors 
$\Delta \beta^{SHG} = \beta_{h}-2\beta_{p} - \beta_{\Lambda}$,  
$\Delta \beta^{OPA}=\beta_{h}-\beta_{s}-\beta_{c}-\beta_{\Lambda}$          
are the phase mismatches of SHG and OPA,
while the relations 
$\kappa_{p}^{SHG}={\kappa_{h}^{SHG}}^{*}$,         
$\kappa_{s}^{OPA}=\kappa_{c}^{OPA}={\kappa_{h}^{OPA}}^{*}$  
can be adopted  from  the Kleinman symmetry.
%

%

%
\subsection{\label{OTR-2-Theoretical-analysis} Theoretical analysis}
%

%
For a theoretical study of cascaded SHG/OPA described in Eqs.~(\ref{Cascaded-SHG/OPA}),
the analytical procedures similar to those used in the previous schemes 
of the two Type-I SHGs are employed.
  In the lossless limit, $\alpha_{i} = 0$,
the complex equations in Eqs.~(\ref{Cascaded-SHG/OPA}) can be formulated again
in terms of real equations to obtain exact theoretical  solutions. 
Along with the complex amplitudes of the electric fields 
in Eqs.~(\ref{parameteru_{psh}}) and
\begin{eqnarray}
E_{c,n}(z) &=& \sqrt{\frac{\omega_{c}}{n_{c} c N_{c}}} u_{c,n}(z) e^{i\phi_{c,n} (z)},
\label{parameteru_{c}(z)}
\end{eqnarray}
 the nonlinear coupling constants can be replaced by 
\begin{eqnarray}
K_{a} &=& 2 \kappa_{a} 
\sqrt{\frac{\omega_{s}}{n_{s}cN_{s}}} 
\sqrt{\frac{\omega_{c}}{n_{c}cN_{c}}} 
\sqrt{\frac{\omega_{h}}{2n_{h}cN_{h}}}
\label{SHG-OPA-K_{a}}
\end{eqnarray}
and $K_{p}$ in Eq.~(\ref{SHG-iSHG-K_{p}}).
For the expressions in Eq.~(\ref{SHG-OPA-K_{a}}), 
the nonlinear coupling constants are defined as
$\kappa_{s}^{OPA} = \kappa_{a} e^{i\delta_{a}}$ 
and the relation
$\kappa_{s}^{OPA} = \kappa_{c}^{OPA} = {\kappa_{h}^{OPA}}^{*}$
is  used. 
Also, another parameter for the nonlinear phase difference is
defined as 
\begin{eqnarray}
\theta_{a,n}(z) &=& \phi_{h,n}(z) - \phi_{s,n}(z) - \phi_{c,n}(z) + \delta_{a}
\label{def-theta_{a,n}}
\end{eqnarray}
 to deal with the phases of the complex electric fields 
and the nonlinear coupling constants.
With all these  considerations, 
the nonlinear equations of the cascaded SHG/OPA 
in Eqs.~(\ref{Cascaded-SHG/OPA}) can be expressed as
\begin{subequations}
  \label{Realtwotype-SHG-OPA}
\begin{eqnarray}
\partial_{z} u_{p,n}(z) &=& -K_{p} u_{p,n}(z) u_{h,n}(z) \sin\theta_{p,n}(z),
\label{SHG-OPA-u_{p,n}}
\\
\partial_{z} u_{s,n}(z) &=& -K_{a} u_{c,n}(z) u_{h,n}(z) \sin\theta_{a,n}(z),
\label{SHG-OPA-u_{s,n}}
\\
\partial_{z} u_{c,n}(z) &=& -K_{a} u_{s,n}(z) u_{h,n}(z) \sin\theta_{a,n}(z),
\label{SHG-OPA-u_{c,n}}
\\
\partial_{z} u_{h,n}(z) &=& K_{p} u_{p,n}^{2}(z) \sin\theta_{p,n}(z)  \nonumber   \\
                        &+& 2 K_{a} u_{s,n}(z) u_{c,n}(z) \sin\theta_{a,n}(z),  
\label{SHG-OPA-u_{h,n}} 
\end{eqnarray}
\begin{eqnarray}
\partial_{z} \theta_{p,n}(z) = \frac{\Gamma_{n}(z)}{u_{h,n}^{2}(z)} 
&+&
 2\frac{\cos\theta_{p,n}(z)}{\sin\theta_{p,n}(z)} \partial_{z} \ln u_{p,n}(z), ~~
\label{SHG-OPA-theta_{p,n}}
\\
\partial_{z} \theta_{a,n}(z) = \frac{\Gamma_{n}(z)}{u_{h,n}^{2}(z)} 
&+&
 \frac{\cos\theta_{a,n}(z)}{\sin\theta_{a,n}(z)} \partial_{z} \ln u_{s,n}(z)   
\nonumber   \\ 
&+&
\frac{\cos\theta_{a,n}(z)}{\sin\theta_{a,n}(z)} \partial_{z} \ln u_{c,n}(z).
\label{SHG-OPA-theta_{a,n}}
\end{eqnarray}
\end{subequations}
For the expressions of 
 Eqs.~(\ref{SHG-OPA-theta_{p,n}})--(\ref{SHG-OPA-theta_{a,n}}), 
an additional  parameter 
$\Gamma_{n}(z)$ is defined as
\begin{eqnarray}
\Gamma_{n}(z) &=&  K_{p} u_{p,n}^{2}(z) u_{h,n}(z) \cos\theta_{p,n}(z)  
\nonumber   \\
              &+& 2K_{a} u_{s,n}(z) u_{c,n}(z) u_{h,n}(z) \cos\theta_{a,n}(z)
 \label{SHG-OPA-Gamma_{n}}
\end{eqnarray}
in the present scheme.
The real equations in Eqs.~(\ref{Realtwotype-SHG-OPA}) are 
the exact equivalents of the complex equations in Eqs.~(\ref{Cascaded-SHG/OPA}). 
The real equations show that the amplitudes of the signal wave 
and the converted idler wave 
are mutually interchanged so as to couple to the second harmonic wave
in the corresponding equations, 
while the parameters of the nonlinear phase difference are coupled with each other 
in the presence of $\Gamma_{n}(z)$.
Then from the nonlinear equations  in Eqs.~(\ref{Realtwotype-SHG-OPA}), 
conserved quantities can be derived for each resonant cycle $n$.
For the first two constants of integration, Manley--Rowe relations 
can be derived as
\begin{subequations}
  \label{SHG-OPA-Manley-Rowe-relations}
\begin{eqnarray} 
 \sum_{i}^{p,s,c,h} u_{i,n}^{2}(z) &=&   \sum_{i}^{p,s,c,h} u_{i,n}^{2}(0),
\label{SHG-OPA-Manley-Rowe-relations-1}
\\
u_{s,n}^{2}(z) - u_{c,n}^{2}(z) &=&  u_{s,n}^{2}(0) - u_{c,n}^{2}(0),
\label{SHG-OPA-Manley-Rowe-relations-2}
\end{eqnarray}
\end{subequations}
from Eqs.~(\ref{SHG-OPA-u_{p,n}})--(\ref{SHG-OPA-u_{h,n}}) and
these relations result in the conservation of total power flow in the waveguide.
 As for the second constant of integration, the parameter $\Gamma_{n}(z)$ is  
an intrinsically conserved quantity expressed as
\begin{eqnarray}
\partial_{z} \Gamma_{n}(z) &=& 0
 \label{SHG-OPA-Gamma_{n}}
\end{eqnarray}
that can be derived from 
Eqs.~(\ref{SHG-OPA-theta_{p,n}})--(\ref{SHG-OPA-theta_{a,n}}). 
 In the case of the conserved quantity $\Gamma_{n}(z)$, if 
  $K_{p} \longrightarrow 0$ or 
  $K_{a} \longrightarrow 0$, 
$\Gamma_{n}(z)$ reduces to the well known constant of integration  
for a single Type-I SHG or OPA in the single pass propagation.
The other two constants of integration, 
which relate the amplitudes of the pump, signal, and idler waves,
 can be derived  from Eqs.~(\ref{SHG-OPA-u_{p,n}})--(\ref{SHG-OPA-u_{c,n}}) 
 and expressed as
 \begin{subequations}
  \label{SHG-OPA-third constant of integration}
\begin{eqnarray}
\frac{\partial_{z} \ln [u_{s,n}(z) + u_{c,n}(z)]}
     {\partial_{z} \ln u_{p,n}(z)} 
&=& 
\frac{K_{a}}{K_{p}} \cdot \frac{\sin\theta_{a,n}(z)}{\sin\theta_{p,n}(z)}, 
 \label{SHG-OPA-third constant of integration-01}
 \\
\frac{\partial_{z} \ln [u_{s,n}(z) - u_{c,n}(z)]}
     {\partial_{z} \ln u_{p,n}(z)} 
&=& 
- \frac{K_{a}}{K_{p}} \cdot \frac{\sin\theta_{a,n}(z)}{\sin\theta_{p,n}(z)}, 
 \label{SHG-OPA-third constant of integration-02}
\end{eqnarray}
\end{subequations}
where the constancy of the relations 
in Eqs.~(\ref{SHG-OPA-third constant of integration})
is the same as that discussed in Sec.~\ref{OTR-1}. 
These relations are associated with the amplitude coupling among the three waves 
and will later be found, in Eqs.~(\ref{SHG-OPA-amplitude-products}), 
to exhibit a correspondence with the Manley–Rowe relation 
in Eq.~(\ref{SHG-OPA-Manley-Rowe-relations-2}).
%

%
\subsection{\label{OTR-2-cSHG/OPA-Solutions} Solutions of cascaded SHG/OPA}
%

\begin{figure}
\includegraphics[width=8.0cm, angle=0]{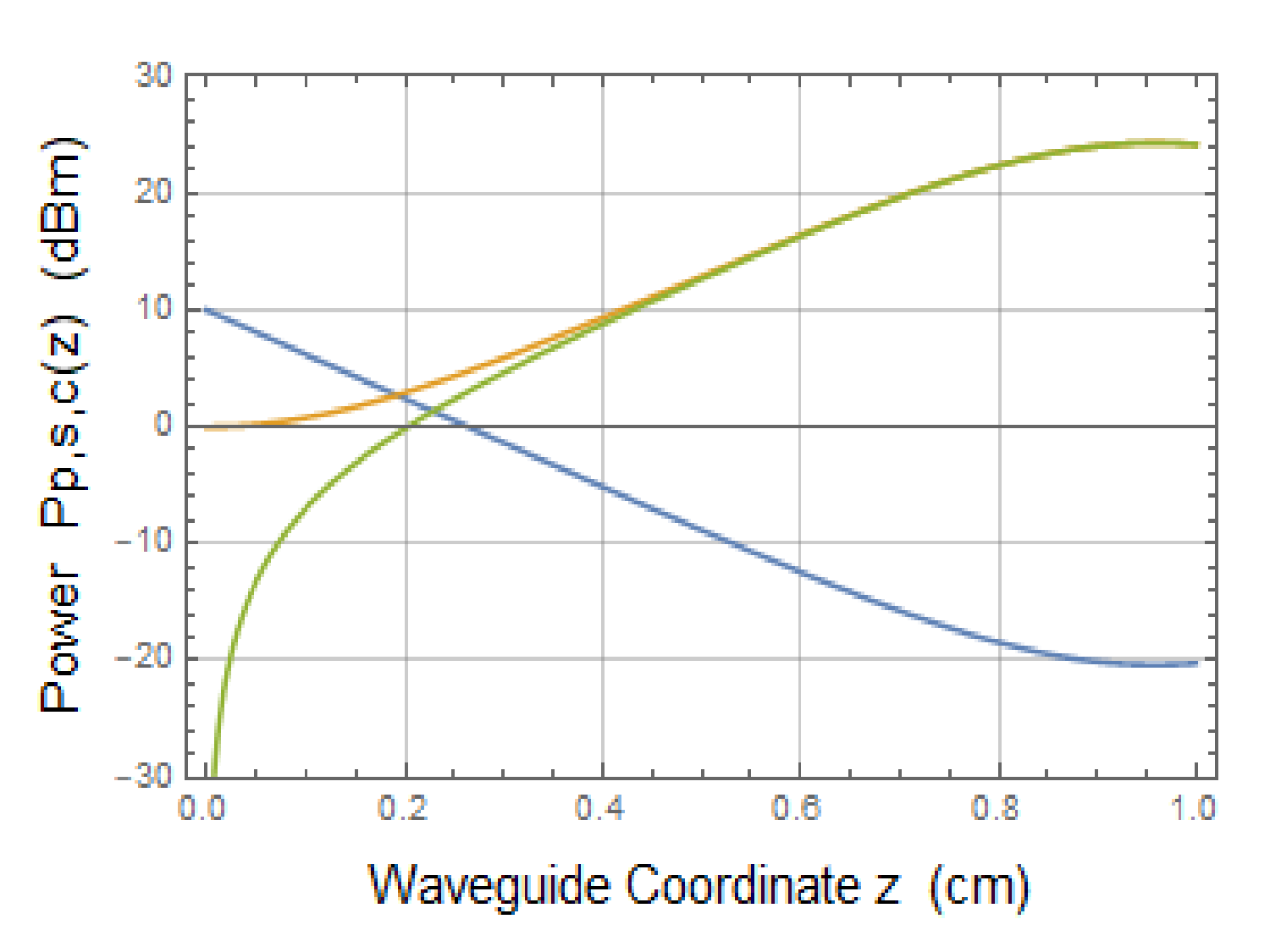}
\caption{
The blue curve shows the deamplification of the pump wave due to SHG, 
while the yellow and green curves show the amplification of the signal wave 
and the generation of the idler wave due to OPA, respectively. 
All curves are plotted on a dBm scale.}
\label{fig:epsart-SHG/OPA-plotting}
\end{figure}

%
The nonlinear equations in Eqs.~(\ref{Realtwotype-SHG-OPA}) are the cascaded 
 processes of Type-I SHG and OPA for the pump, the signal, 
 and the converted idler wave. 
The equations involve  six parameters for the four amplitudes and the two
nonlinear phase differences to be solved.
The second constant of integration in  Eq.~(\ref{SHG-OPA-Gamma_{n}})  
provides the solutions for the two parameters of the nonlinear phase difference in Eqs.~(\ref{SHG-OPA-theta_{p,n}})--(\ref{SHG-OPA-theta_{a,n}})  
   which  in turn can be used to  
   solve the four equations in Eqs.~(\ref{SHG-OPA-u_{p,n}})--(\ref{SHG-OPA-u_{h,n}}).
As discussed in Sec.~\ref{OTR-1}, 
the absence of the initial input for the second harmonic wave 
 in Eq.~(\ref{u_{h,0}(0) = 0})  leads to
$ \Gamma_{n}(z) = 0$
in the present cascaded SHG/OPA as well.
 Consequently, the solutions of the nonlinear equations 
 in Eqs.~(\ref{SHG-OPA-theta_{p,n}})--(\ref{SHG-OPA-theta_{a,n}}) are
\begin{eqnarray}
 \theta_{p,n}(z) = \pm  \frac{\pi}{2}, ~ 
 \theta_{a,n}(z) =  \pm  \frac{\pi}{2}
\label{SHG-OPA-theta_{p,n}(z)-theta_{a,n}(z)-solutions}
\end{eqnarray}
  from $\cos\theta_{p,n}(z) = \cos\theta_{a,n}(z) = 0$.
The nonlinear phase differences are constant and  conserved 
in the nonlinear resonant structure. 
In the present scheme, if  $K_{p} = K_{a}$ is assumed
with $\sin\theta_{p,n}(0) = - \sin\theta_{a,n}(0)$ 
and under the approximations
$1 - \omega_{s} \omega_{c}/\omega_{p}^{2}  
= (\Delta \omega)^{2}/\omega_{p}^{2} \ll 1$,
$1 - n_{s} n_{c}/n_{p}^{2}  
= (\Delta n)^{2}/n_{p}^{2} \ll 1$, 
then
the third constants of integration 
in Eqs.~(\ref{SHG-OPA-third constant of integration}) reduce to
\begin{subequations}
  \label{SHG-OPA-third-relation-of integration}
\begin{eqnarray}
\frac{\partial_{z} \ln [u_{s,n}(z) + u_{c,n}(z)]}
     {\partial_{z} \ln u_{p,n}(z)}   &=& -1, 
 \label{SHG-OPA-third-relation-of-integration-1}
 \\
\frac{\partial_{z} \ln [u_{s,n}(z) - u_{c,n}(z)]}
     {\partial_{z} \ln u_{p,n}(z)}   &=& 1, 
 \label{SHG-OPA-third-relation-of-integration+1}
\end{eqnarray}
\end{subequations}
and yield relations among the amplitudes
 of the pump,  signal, and  idler waves as shown below
\begin{subequations}
  \label{SHG-OPA-amplitude-products}
\begin{eqnarray}
u_{s,n}(z) + u_{c,n}(z) = [u_{s,n}(0) + u_{c,n}(0)] 
\frac{u_{p,n}(0)}{u_{p,n}(z)},  ~ 
\label{SHG-OPA-amplitude-product-1}
\\
u_{s,n}(z) - u_{c,n}(z) = [u_{s,n}(0) - u_{c,n}(0)] 
\frac{u_{p,n}(z)}{u_{p,n}(0)}. ~ 
\label{SHG-OPA-amplitude-product+1}
\end{eqnarray}
\end{subequations}
The relations in Eqs.~(\ref{SHG-OPA-amplitude-products}) show that 
the summation of the amplitudes of the signal wave and the idler wave
is inversely proportional to the pump wave amplitude,
while 
the difference between the two amplitudes 
is linearly proportional to the pump wave amplitude.
Therefore the relations indicate that, 
if the pump wave is deamplified,
 the signal wave and the idler wave can be co-amplified 
whereas the difference between the two amplitudes  can be reduced. 
 After some algebra similar to that used in the schemes in Sec.~\ref{OTR-1}, 
 analytical solutions of the cascaded SHG/OPA processes can be obtained from 
 Eqs.~(\ref{SHG-OPA-u_{p,n}})--(\ref{SHG-OPA-u_{h,n}})  as 
\begin{widetext}
\begin{subequations}
  \label{solutions-of-SHG/OPA}
\begin{eqnarray}
u_{p,n}(z)  &=& \frac{U_{n}(0)}{\sqrt{2} V_{n}(0) \sqrt{1+k_{n}^{2}}} 
[\textnormal{dn} \{b(z+z_{n}), k_{n}\} + k_{n} \textnormal{cn} \{b (z+z_{n}), k_{n}\}],
\label{solutions-of-SHG/OPA-u_{p,n}}
\\ 
u_{s,n}(z) &=& \frac{1}{2}  \frac{U_{n}(0)}{\sqrt{1+k_{n}^{2}}} 
[\textnormal{dn} \{b(z+z_{n}), k_{n}\} - k_{n} \textnormal{cn} \{b (z+z_{n}), k_{n}\}]
\nonumber \\ 
&+& \frac{1}{2}  \frac{\sqrt{V_{n}^{2}(0) - 1}}{V_{n}(0)} 
 \frac{U_{n}(0)}{\sqrt{1+k_{n}^{2}}}  
[\textnormal{dn} \{b(z+z_{n}), k_{n}\} + k_{n} \textnormal{cn} \{b (z+z_{n}), k_{n}\}],   
\label{solutions-of-SHG/OPA-u_{s,n}}
\\
u_{c,n}(z) &=& \frac{1}{2}  \frac{U_{n}(0)}{\sqrt{1+k_{n}^{2}}} 
[\textnormal{dn} \{b(z+z_{n}), k_{n}\} - k_{n} \textnormal{cn} \{b (z+z_{n}), k_{n}\}]
\nonumber \\ 
&-& \frac{1}{2}  \frac{\sqrt{V_{n}^{2}(0) - 1}}{V_{n}(0)} 
 \frac{U_{n}(0)}{\sqrt{1+k_{n}^{2}}}  
[\textnormal{dn} \{b(z+z_{n}), k_{n}\} + k_{n} \textnormal{cn} \{b (z+z_{n}), k_{n}\}],   
\label{solutions-of-SHG/OPA-u_{c,n}}
\\
u_{h,n}(z) &=& \frac{\sqrt{2} k_{n} U_{n}(0)}{\sqrt{1+k_{n}^{2}}}  
\textnormal{sn} [b(z+z_{n}), k_{n}].
\label{solutions-of-SHG/OPA-u_{h,n}}
\end{eqnarray}
\end{subequations}
\end{widetext}
For the simplified expressions of the solutions, 
six constants $U_{n}(0)$, $V_{n}(0)$, $W_{n}(0)$, $k_{n}$, $b$,
 and $z_{n}$ are introduced. 
The expressions for the three constants $U_{n}(0)$, $V_{n}(0)$, and $W_{n}(0)$
relevant to the amplitudes can be determined from
%
%
\begin{subequations}
  \label{SHG-OPA-U_{n}(0)-V_{n}(0)-W_{n}(0)}
\begin{eqnarray}
U_{n}^{2}(0) 
&=& u_{p,n}^{2}(0) + u_{s,n}^{2}(0) + u_{c,n}^{2}(0) + u_{h,n}^{2}(0),
\label{SHG/OPA-U_{n}(0)} ~~ 
\\
V_{n}^{2}(0) &=& 1 +  \frac{1}{2} 
                      \frac{[u_{s,n}(0) - u_{c,n}(0)]^{2}}{u_{p,n}^{2}(0)},
\label{SHG/OPA-V_{n}(0)}  
\\
W_{n}^{2}(0) &=& \frac{1}{2} u_{p,n}^{2}(0)[u_{s,n}(0) + u_{c,n}(0)]^{2},
\label{SHG/OPA-W_{n}(0)} 
\end{eqnarray}
\end{subequations}
%
and the parameters $k_{n}$, $b$, and  $z_{n}$ 
can be defined  as
\begin{eqnarray}
k_{n}^{2} &=& \frac{U_{n}^{2}(0) - 2V_{n}(0) W_{n}(0)}{U_{n}^{2}(0) + 2V_{n}(0) W_{n}(0)}
\label{SHG/OPA-k_{n}},    \\
%
b &=& \frac{\sqrt{2} K_{p} \sin\theta_{p,n}(0) U_{n}(0)}{\sqrt{1+k_{n}^{2}}},
\label{SHG/OPA-b}     \\
z_{n} &=& \frac{1} {b} \textnormal{sc}^{-1} 
[\frac{\sqrt{2} u_{h,n}(0)} 
{\sqrt{2} V_{n}(0) u_{p,n}(0) - u_{s,n}(0) - u_{c,n}(0)}],
  \label{SHG/OPA-z_{n}} ~~  
%
\end{eqnarray}
in the present scheme.
The representations  in Eq.~(\ref{SHG-OPA-theta_{p,n}(z)-theta_{a,n}(z)-solutions}) 
and Eqs.~(\ref{solutions-of-SHG/OPA}) 
are the exact solutions 
for the cascaded SHG/OPA with $K_{p} = K_{a}$  in Eqs.~(\ref{Realtwotype-SHG-OPA}). 
The solutions in Eqs.~(\ref{solutions-of-SHG/OPA}) include
the cascaded SHG/OPA interactions that lead to the simultaneous amplification of 
 the signal wave and the idler wave 
in the range $0 \leq z \leq L$ of the nonlinear resonant structure.
In the limit as $k_{n} \rightarrow 0$ in the solutions,  
a condition 
\begin{eqnarray} 
u_{p,n}^{2}(0) = 2u_{s,n}(0)u_{c,n}(0), ~ u_{h,n}(0) = 0
\label{SHG/OPA-transparency condition}
\end{eqnarray}
  is  derived from 
  Eqs.~(\ref{SHG-OPA-U_{n}(0)-V_{n}(0)-W_{n}(0)})--(\ref{SHG/OPA-k_{n}})
  and under this limit, the condition leads to the solutions
\begin{eqnarray} 
u_{i,n}(z) &=& u_{i,n}(0), ~  (i=p,s,c,h)
\label{SHG/OPA-transparent-solutions}
\end{eqnarray}
 from  Eqs.~(\ref{solutions-of-SHG/OPA}).
 The solutions in Eq.~(\ref{SHG/OPA-transparent-solutions}) show  
 the phenomenon of nonlinear transparency
 that illustrates 
 the independent transparent propagation
of the pump, the signal, and the idler waves
without the nonlinear interactions arising from the balance and the equilibrium 
established between SHG and OPA.
The SHG/OPA transparency can be demonstrated analytically
from Eqs.~(\ref{SHG-OPA-u_{p,n}})--(\ref{SHG-OPA-u_{h,n}})
and can also be confirmed
 from the numerical calculations  of the complex equations 
in Eqs.~(\ref{Cascaded-SHG/OPA}) with the initial conditions of the four waves,
  for example $P_{p,n}(0) = 10$ mW, $P_{s,n}(0) = P_{c,n}(0) = 5$ mW, and
  $P_{h,n}(0) =0$, 
as required  in Eq.~(\ref{SHG/OPA-transparency condition}).
In the limit as $k_{n} \rightarrow 1$,  
 a condition  $u_{s,n}(0) + u_{c,n}(0) = 0$ 
 is derived  from Eq.~(\ref{SHG/OPA-k_{n}}) 
 and  the solutions in Eqs.~(\ref{solutions-of-SHG/OPA}) reduce to
%
\begin{subequations}
  \label{solutions-of-SHG/OPA-k_{n}=1}
\begin{eqnarray}
u_{p,n}(z) &=& 
 \frac{u_{p,n}(0) U_{n}(0) \textnormal{sech} [b (z+z_{n})]} 
      {\sqrt{u_{p,n}^{2}(0) + u_{s,n}^{2}(0) + u_{c,n}^{2}(0)}},
\label{solutions-of-SHG/OPA-k_{n}=1-u_{p,n}}
\\
u_{s,n}(z) &=& 
 \frac{u_{s,n}(0) U_{n}(0) \textnormal{sech} [b (z+z_{n})]} 
      {\sqrt{u_{p,n}^{2}(0) + u_{s,n}^{2}(0) + u_{c,n}^{2}(0)}},
\label{solutions-of-SHG/OPA-k_{n}=1-u_{s,n}}
\\
u_{c,n}(z) &=& 
 \frac{u_{c,n}(0) U_{n}(0) \textnormal{sech} [b (z+z_{n})]} 
      {\sqrt{u_{p,n}^{2}(0) + u_{s,n}^{2}(0) + u_{c,n}^{2}(0)}},
\label{solutions-of-SHG/OPA-k_{n}=1-u_{c,n}}
\\
u_{h,n}(z) &=& U_{n}(0) \textnormal{tanh} [b (z+z_{n})].
\label{solutions-of-SHG/OPA-k_{n}=1-u_{h,n}}
\end{eqnarray}
\end{subequations}
%
The solutions in Eqs.~(\ref{solutions-of-SHG/OPA-k_{n}=1}) include
SHG and sum frequency generation 
for the amplification of the second harmonic wave
via the deamplifications of the pump,  the signal, and the idler waves.
In the limit $u_{p,n}(0) \rightarrow 0$ or $u_{s,n}(0)$, $u_{c,n}(0) \rightarrow 0$, 
 the solutions  reduce to the well known solutions 
 of the corresponding nonlinear interactions in the case of $k_{n} = 1$.
Numerical solutions in the scheme of the cascaded SHG/OPA
can be provided from the direct calculations of  Eqs.~(\ref{Cascaded-SHG/OPA})
under  the condition in Eq.~(\ref{resonant-feedback-E_{h,n+1}(0)}). 
For the  convenience of calculations, 
let us assume that the nonlinear coupling constant
for the cascaded SHG/OPA is $d_{33}$ and the Type-0 quasi-phase matching.
All the parameters in the  cascaded SHG/iSHG calculations are
used  but   
$d_{33} =30$ pm/V,
 $\alpha_{h} = 3\alpha_{p} = 0.03$ dB/cm ,
 $\lambda_{s} = 1551$ nm, and
$L = 1$ cm 
are adopted.
Together with  the input $P_{c,n}(0) = 0$, 
the power of the three waves $P_{p,s,c}(z)$ from the solutions 
is plotted 
as a function of the waveguide coordinate $z$ 
 in Fig.~\ref{fig:epsart-SHG/OPA-plotting} on a dBm scale.
%
%
In the linear scale version of Fig.~\ref{fig:epsart-SHG/OPA-plotting}, 
the signal and idler traces are not visually distinguishable 
due to both the nearly equal gain of the two waves and the chosen plotting scale. 
This nearly equal gain is a well known property of OPA 
that follows directly from the Manley--Rowe relation 
in Eq.~(\ref{SHG-OPA-Manley-Rowe-relations-2}). 
Moreover, because the amplified peak powers 
of both waves reach $\sim 250~\mathrm{mW}$ 
while their difference remains nearly constant 
at $\sim 1~\mathrm{mW}$ (set by the initial conditions), 
the linear y-axis scale causes the two traces to overlap.
%
%

%
The deamplification of the pump wave due to SHG 
that defines the digital off state
and the amplification of the signal and the idler wave due to OPA 
that defines the digital on state
are demonstrated in the figure. 
From the numerical results of the collector power 
 $P_{s,n}(L) + P_{c,n}(L) = 522.6$ mW,
 the power transfer ratio $\alpha_{TR} = 52.26$
 and  
 the power amplification factor $\beta_{AF} = 522.6$
 can be obtained. 
When compared with the results of the cascaded SHG/iSHG,
the waveguide  with the large nonlinear coupling constant 
and the small propagation loss constant \cite{Shams-Ansari2022} 
is necessary
for the implementation of the efficient optical transistor. 
%
%
For a waveguide of length $L$,
the response time of the present device 
can be defined as the propagation time of the signal wave 
through the resonator,
\begin{equation}
\Delta t_{\text{res}} = \frac{n_s}{c} \cdot L
\label{response time for L}
\end{equation}
which, for $n_s = 2.1544$ and $L = 1$~cm, yields 
$\Delta t_{\text{res}} \simeq 71.9$~ps, 
much shorter than the reported response time of
LiNbO$_3$ optical switches ($<10$~ns) \cite{Kuratani2010CLEO}.
%

%
For comparison with the reports in Refs.~[1--13], 
in the single frequency configuration 
with $-d_{21} = d_{22} = 3$ pm/V and 
     $\alpha_{h} = 3\alpha_{p} = 0.003$ dB/cm,
the performance reaches $\alpha_{TR}=4.838$ and $\beta_{AF}=48.38$, 
whereas in the dual frequency configuration 
with $d_{33} =$ 30 pm/V and
     $\alpha_{h} = 3\alpha_{p} = 0.03$ dB/cm,
the corresponding values increase to $\alpha_{TR}=52.26$ and $\beta_{AF}=522.6$. 
These results highlight the specific performance improvements 
of the present structure compared to existing models, 
while also clarifying the relative merits and limitations 
between the single and dual frequency implementations.
%
With respect to sensitivity analyses,
the nonlinear coefficient $d_{il}$ 
and the propagation loss constant $\alpha_i$ are dominant factors 
and thus can serve as useful figures of merit.
Variations in these parameters 
lead to significant changes in $\alpha_{TR}$ and $\beta_{AF}$,
thereby 
influencing the performance of the optical transistor,
as discussed in the representative cases.
%
To enhance $\alpha_{TR}$ and $\beta_{AF}$ in the single frequency  transistor, 
it is indispensable to increase the nonlinear coefficients 
$d_{21}$ and $d_{22}$, 
which constitutes an intrinsic limitation to be overcome 
through materials research.
%

%
The proposed optical transistor is expected to serve as a key optical device 
for amplifiers, switches, logic gates, and integrated circuits, 
with broad applications in all optical signal processing, 
particularly in communication and computing. 
It offers the potential for high-speed operation
while maintaining low power consumption,
thereby addressing the performance limitations
of conventional electronic transistors.
The present device can also be applied to other frequency regimes, 
such as the terahertz range, 
provided that a nonlinear resonant structure 
with comparable properties can be realized.
As for future research, the proposed optical transistor is 
timely for experimental realization,
as suitable materials and technologies are already available 
and supported by a solid theoretical foundation. 
A Fabry--Perot microcavity platform may appear to be a promising candidate; 
for example, a thin-film lithium-niobate-on-insulator microcavity 
based on a strip-loaded waveguide and distributed Bragg reflector mirrors 
in Ref.~\cite{Kotlyar2021} may require only periodic poling 
for the dual frequency design, 
whereas the single frequency design would additionally require advances 
in waveguide technology.
%
%

%
\section{\label{Conclusions} Conclusions}
%
%
The present study theoretically establishes an optical transistor 
based on cascaded second-order nonlinear interactions 
within a resonant waveguide structure. 
Two representative schemes are analyzed: 
the cascaded SHG/iSHG scheme operating with waves of a single frequency 
and the cascaded SHG/OPA scheme operating with dual frequencies. 
When only the emitter wave is incident, 
its second harmonic wave is generated and resonates inside the structure, 
resulting in the off state of the collector output.
When the base wave is applied together with the emitter wave, 
the collector wave appears in the on state,
so that the base wave controls the emitter wave to manipulate amplification 
and on–off switching of the collector output.
Exact theoretical solutions supported by numerical calculations
reveal that both schemes realize transistor-like operation characterized by 
simultaneous amplification and switching that define the digital on/off states.
In both schemes, 
the phenomenon of nonlinear transparency is theoretically predicted 
and numerically confirmed, illustrating the transparent propagation 
of the input waves in the absence of nonlinear coupling.

In the single frequency configuration, 
cascaded SHG/iSHG enables cascadable signal amplification 
and satisfies the fan-out criterion with moderate power requirements. 
In the dual frequency configuration, 
cascaded SHG/OPA achieves substantially higher amplification 
and transfer efficiency 
because of the intrinsically larger nonlinear coefficient of LiNbO$_3$.
Both configurations satisfy the fundamental criteria 
for optical transistor operation, 
including cascadability and fan-out, confirming their functional viability.
Furthermore, the achieved metrics of power transfer ratio 
and power amplification factor indicate performance levels 
comparable to those of electronic transistors.
The theoretical framework developed in this work provides a consistent formulation
for describing the fundamental operating principles of optical transistors
based on a nonlinear resonant structure.
These findings demonstrate that an all optical transistor can be realized 
as a single compact device operating at high speed 
and low power consumption, suggesting its applicability 
in integrated photonic circuits for optical communication and computing.
%
%

\begin{acknowledgments}
This work was supported by  projects under
 Grant Nos. ETRI 25ZS1200, K25L5M1C1,
and by the Professional Experienced Personnel Invitation Program (2024-1-0149),
all funded by the MSIT of Korea.
\end{acknowledgments}
%
\bibliography{OTR-arXiv-Jongbae-bib-01-15-26}

\end{document}